\documentclass[10pt,letterpaper]{article}
\usepackage[top=0.85in,left=2.75in,footskip=0.75in]{geometry}
\usepackage{amsmath,amssymb}

\usepackage{changepage}

\usepackage{textcomp,marvosym}

\usepackage{cite}

\usepackage{nameref,hyperref}

\usepackage[right]{lineno}

\usepackage{microtype}
\DisableLigatures[f]{encoding = *, family = * }

\usepackage[table]{xcolor}

\usepackage{array}

\newcolumntype{+}{!{\vrule width 2pt}}

\newlength\savedwidth

\raggedright
\usepackage[aboveskip=1pt,labelfont=bf,labelsep=period,justification=raggedright,singlelinecheck=off]{caption}

\makeatletter
\renewcommand{\@biblabel}[1]{\quad#1.}
\makeatother

\usepackage{lastpage,fancyhdr,graphicx}
\usepackage{epstopdf}
\fancyheadoffset[L]{2.25in}
\fancyfootoffset[L]{2.25in}
\usepackage[capitalize]{cleveref}

\newcommand\eps{\ensuremath{\varepsilon}}

\DeclareMathOperator*{\argmax}{arg\,max}

\newcommand{\new}[1]{{{#1}}}

\def\kp{k}
\def\kd{\kappa}

\def\betaGhost{}
\def\betagGhost{\beta_g}
\def\Ug{V}
\def\ug{v}

\usepackage{color}

\def\gsim{\mathrel{\raise.3ex\hbox{$>$\kern-.75em\lower1ex\hbox{$\sim$}}}}
\def\lsim{\mathrel{\raise.3ex\hbox{$<$\kern-.75em\lower1ex\hbox{$\sim$}}}}

\usepackage{ulem}

\makeatletter
\let\saved@includegraphics\includegraphics
\AtBeginDocument{\let\includegraphics\saved@includegraphics}
\renewenvironment*{figure}{\@float{figure}}{\end@float}
\makeatother

\begin{document}
\vspace*{0.2in}

\begin{flushleft}
	{\Large
		\textbf\newline{Rational social distancing policy during epidemics with limited healthcare capacity} 
	}
	\newline
	\\
	Simon K. Schnyder\textsuperscript{1*},
	John J. Molina\textsuperscript{2},
	Ryoichi Yamamoto\textsuperscript{2},
	Matthew S. Turner\textsuperscript{3,4*}
	\\
	\bigskip
	\textbf{1} Institute of Industrial Science, The University of Tokyo, 4-6-1 Komaba, Meguro, Tokyo, Japan
	\\
	\textbf{2} Department of Chemical Engineering, Kyoto University, Kyoto, Japan
	\\
	\textbf{3} Department of Physics, University of Warwick, Coventry, UK
	\\
	\textbf{4} Institute for Global Pandemic Planning, University of Warwick, Coventry, UK
	\\
	\bigskip

	%
	%





	* To whom correspondence should be addressed. SKS: skschnyder@gmail.com, MST: m.s.turner@warwick.ac.uk.

\end{flushleft}
\section*{Abstract} 

Epidemics of infectious diseases posing a serious risk to human health have occurred throughout history. During recent epidemics there has been much debate about policy, including how and when to impose restrictions on behaviour. Policymakers must balance a complex spectrum of objectives, suggesting a need for quantitative tools. Whether health services might be `overwhelmed' has emerged as a key consideration. Here we show how costly interventions, such as taxes or subsidies on behaviour, can be used to exactly align individuals' decision making with government preferences even when these are not aligned.
In order to achieve this, we develop a nested optimisation algorithm of both the government intervention strategy and the resulting equilibrium behaviour of individuals.
We focus on a situation in which the capacity of the healthcare system to treat patients is limited and identify conditions under which the disease dynamics respect the capacity limit. We find an extremely sharp drop in peak infections  at a critical maximum infection cost in the government's objective function. This is in marked contrast to the gradual reduction of infections if individuals make decisions without government intervention. We find optimal interventions vary less strongly in time when interventions are costly to the government and that the critical cost of the policy switch depends on how costly interventions are.

\section*{Author summary}
The question of how to determine policies during epidemics is a subject of broad contemporary interest. How and when should society impose behavioural restrictions in order to reduce infections? Policymakers have to balance many objectives, suggesting a need for quantitative tools for designing optimal intervention policy.  Previous work on optimal policy-making typically sidesteps the question of how the population follows any intervention chosen by a government. Furthermore, the cost of implementing an intervention is also usually ignored. Our work overcomes these shortcomings. We analyse how the population chooses behaviour in a self-organised way. This can be influenced by the government so as to optimise its own objective function. Its objectives can be different from the individuals. Our work represents a proof-of-principle that costly policy interventions can be developed in the environment of (i) a disease with well-understood epidemiological character (ii) health-care capacity limits (iii) where those interventions are costly to implement. For these reasons we believe it highlights an opportunity to develop deployable policymaking tools and further advances our understanding of epidemiology when individuals adapt their behaviour in response to perceived dangers.


\section*{Introduction}

Policymakers can manage epidemics using a variety of non-clinical interventions that target behaviour and hence the rate at which the disease is passed on. At one extreme this can involve merely providing accurate information and/or conceptual tools to enable rational individuals to identify their optimum behaviour. More interventionist strategies available to policymakers include subsidising preferred behaviour and/or penalising behaviour that they wish to discourage.  Recent epidemics have generated much debate about policy, including how and when to impose restrictions on behaviour. Policy is likely to fall sharply into focus as the epidemic is analysed in a historical context, informing our planning for future epidemics.
The primary goal of this work is to establish a proof-of-principle that fully quantitative approaches can be used to help design optimal intervention strategies, first in a stylised model but without obvious conceptual limits to incorporating more faithful descriptions of population composition and behaviour. We show that the costs of government interventions can be incorporated into the kinds of quantitative tools that would be necessary to manage future pandemics.

Dealing with an epidemic as a policymaker requires a number of objectives to be prioritised and balanced. The goal of limiting infections may justify restrictions on the day-to-day social and economic activities of citizens or subjects. A {\it rational} policy design process involves policymakers who are aware of the strategies that provide the most beneficial outcomes, these being evaluated using quantitative metrics.
Our motivation here is to further the development of such quantitative tools. Ultimately we would see this as an aid to policy making but here we are concerned with establishing a point of principle - that it is possible to target outcomes that are optimal in the sense that they maximise an objective function that balances costs against benefits in the specific case of (i) when these interventions can carry costs to the government, (ii) when the healthcare system has limited capacity, (iii) when the interventions (have to) take into account the endogenous behaviour of the population, including its response to said interventions, and (iv) when the interests of government and population might not be aligned.

This study is concerned with rational policymaking in and for a society of rational individuals. There already exists a literature that explores the behaviour of rational {\it individuals}, in the absence of policy interventions. These individuals are typically assumed to be able to adjust their behaviour in the face of an epidemic \cite{Reluga2010,Fenichel2011,Wang2016,Chang2020,Verelst2016,Bhattacharyya2019,Makris2020,Yan2021,Reluga2011}.
Broadly speaking, individuals may choose to limit their social activity when infections are high, to avoid the risk of becoming infected themselves, provided that the health risks outweigh the social and economic costs.
In the opposite limit, little or no behavioural changes are made and the epidemic is assumed to run its natural course much as if the agents were unreasoning. These studies are highly stylised in several respects, including the use of population-wide mean-field compartmentalised model and little or no analysis of the role of uncertainties. While they have not yet been developed into the more sophisticated variants needed to reorientate towards real data they nonetheless lay down an important milestone in demonstrating that such analysis is possible, at least in principle. It is generally straightforward to see how such approaches can be extended to incorporate the complexities of real data, mirroring the sophistication of epidemiological approaches that incorporate more realistic household-level descriptions. This might include multiple compartment types with different risk and behaviour profiles \cite{Acemoglu2020,Fenichel2011,Prem2017,Huang2022,Tildesley2022},
spatial \cite{Chandrasekhar2021} and temporal networks \cite{Holme2012,Holme2015}, seasonal effects \cite{He2013}, spatial or transmission heterogeneity~\cite{Mossong2008,Tildesley2010,Prem2017,Sun2021}
or agent-based models \cite{Ferguson2006,Tanimoto2018,Mellacher2020,Grauer2020}.
It is also possible to include noise, for instance in the control \cite{yong1999stochastic}.
{It is also of interest to study the inverse problem to ours where one attempts to infer the objective function underlying some observed (social distancing) behaviour~\cite{Molina2022}.}

Perhaps the most fundamental common assumption is that individual agents act rationally, i.e. to maximise an economic utility. Although the limitations of such approaches have been widely acknowledged, e.g. within behavioural economics generally \cite{kahneman2003maps}, this remains one of the fundamental assumptions of modern economic theory and will be adopted in the present work, noting that conceptual tools could be provided to assist individuals in identifying rational decisions. Recent methodological advances have allowed to establish the behaviour of individuals that target a Nash equilibrium, rather than a global utility maximum that requires coordination \cite{Reluga2010,Reluga2011,Fenichel2011,McAdams2020,Eichenbaum2021}.

Different from such \textit{decentralised} decision-making, governments present an instance of \textit{centralised} decision-making. These will typically not aim for Nash equilibria but for policy that is more socially optimal or better aligned with political or national priorities~\cite{Rowthorn2020,Toxvaerd2020,Makris2020,Li2017}.
Furthermore, subsidy and tax schemes can be used by a social planner to \textit{decentralize optimal policy}, i.e. to bring the Nash equilibrium of individuals into alignment with the global optimum \cite{Rowthorn2020,Bethune2020,Aurell2022}.
These approaches have so far only been applied to the special case where the subsidy and tax schemes are cost-free to the social planner, i.e. they appear only in the utility function of the individuals. Additionally, attention has been restricted to the case where the preferences of the government and the population are well aligned. We go beyond these restrictions by invoking a hierarchy of interests. This requires a nested optimisation of both the government intervention strategy and the underlying equilibrium behaviour of the population. An important aspect of our work is that we investigate the situation in which the cost of an infection relative to the cost of social distancing can be quite different for the government than for an individual. This is highly plausible, as for instance, it is likely to be more difficult for an individual to negotiate the right to work remotely than were the government to impose these arrangements.

Typical government interventions in this literature would involve taxing high social activity of infectious individuals {with the aim of disincentivising}  them from certain behaviour, akin to a Pigouvian tax or subsidy \cite{Rowthorn2020,Althouse2010}. The collected taxes get redistributed equally over the whole population. The typical assumption is then that the intervention has no direct effect on the government's objective function since the \textit{process} of redistribution is assumed to be cost-free, while of course the results of the taxation, here the reduction of social activity, do impact the government objective function indirectly. However, we argue that one must consider the \textit{process} of redistribution itself as costly.
e.g. due to
the misallocation of resources and the distortion of markets caused by the collection of taxes, an effect known as the shadow or marginal cost of public funds \cite{Stiglitz1971}.
{Another factor could be that the administration of the incentivisation process is in itself costly, e.g. it requires clerical and professional resources, surveillance resources, etc.}

Some recent studies have also focussed on the role of healthcare
thresholds \cite{Palmer2020,Hayhoe2020,Piguillem2020,Kantner2020,Eichenbaum2021,Ketcheson2021}, but not in combination with Nash equilibrium behaviour {and} costly government interventions.
Ref. \cite{Eichenbaum2021}
is most similar to ours, investigating the role of government intervention on equilibrium behaviour in a situation where the case fatality rate depends on the current number of cases. Their work differs from ours in that they study equations that are discrete in time, and that their case fatality rate is unbounded for large infection numbers. Most importantly, they are only interested in the case in which the government and the individuals have the same preferences and that the intervention is cost-free.

SIR models being compartmental models with continuous values, it is impossible to fully eradicate the disease, at best reaching an exponential decay of infections with strong social distancing or after reaching herd immunity via infections or vaccination. While eradication can in principle be incorporated, e.g. by defining a critical value of the infectious compartment below which the disease is said to have been eradicated \cite{Piguillem2020}, eradication is quite complicated to reach in a global pandemic in practice. This is why we choose to neglect the possibility of complete eradication in what follows.

Waves of infections are predicted to occur under certain circumstances, e.g. when fresh variants occur that (partially) escape immunity \cite{Schwarzendahl2022}, waning immunity and demographics \cite{Giannitsarou2021}, or when social distancing is a more {\it ad hoc} response to recent changes in the infection and fatality numbers \cite{Lux2021}.

We focus on calculating the self-organised social distancing of individuals and the government incentives that enable such behaviour. We do not investigate other possible policy interventions such as vaccination and treatment strategies, \cite{Bauch2003,Bauch2004,Reluga2006,Tildesley2006,Reluga2011,Chen2014,Wang2016,Tanimoto2018,Chang2020,Toxvaerd2020,Grauer2020,Moore2021b,Moore2022,Hill2022,Keeling2023}, or isolation, testing, and active case-tracing strategies \cite{Kucharski2020,Piguillem2020}, noting that these can be included in future variants of models like the one we analyse here.
{Instead, we assume that a vaccine becomes available at a time far longer than the duration of the epidemic, at which point all the remaining susceptible people become immune to the disease instantly. We do this so that we only have to study the behaviour on a finite time horizon. We ignore the situation where a vaccine becomes available during the epidemic.} {While the early arrival of a vaccine would have consequences for both equilibrium and globally optimal behaviour \cite{Reluga2010,Makris2020, Eichenbaum2021,Schnyder2023}, this lies outside of the scope of this work. Judging from previous work, one would roughly expect that the earlier the vaccine is expected to arrive, the more incentivised both individuals and governments would be to  increase their social distancing efforts.}

In what follows policymakers are also assumed to be acting rationally. They decide how to intervene so as to maximise a government-level objective function. In the spirit of a proof of principle we limit policy priorities to three of the most obvious factors: reducing direct health risks, avoiding excessive stress on the health care system and mitigating the social and financial impact associated with placing limits on individual behaviour. The primary variables are: (1) the  infectiousness $k(t)$, parameterising the mean number of additional cases a single infected individual would cause in a previously unexposed population. This is assumed to have a background, or natural, level $\kappa^*>1$ adopted by society in the absence of any behavioural changes, also known as the basic reproduction number $R_0$. (2) A time-dependent government intervention $\eps(t)$ that can be deployed to incentivise behavioural changes in individuals. For simplicity we neglect the possibility of reinfection, although the present framework can be modified to incorporate this.

\section*{Methods}

\begin{figure}[tbp]
	\centering
	\includegraphics[width=0.65\columnwidth]{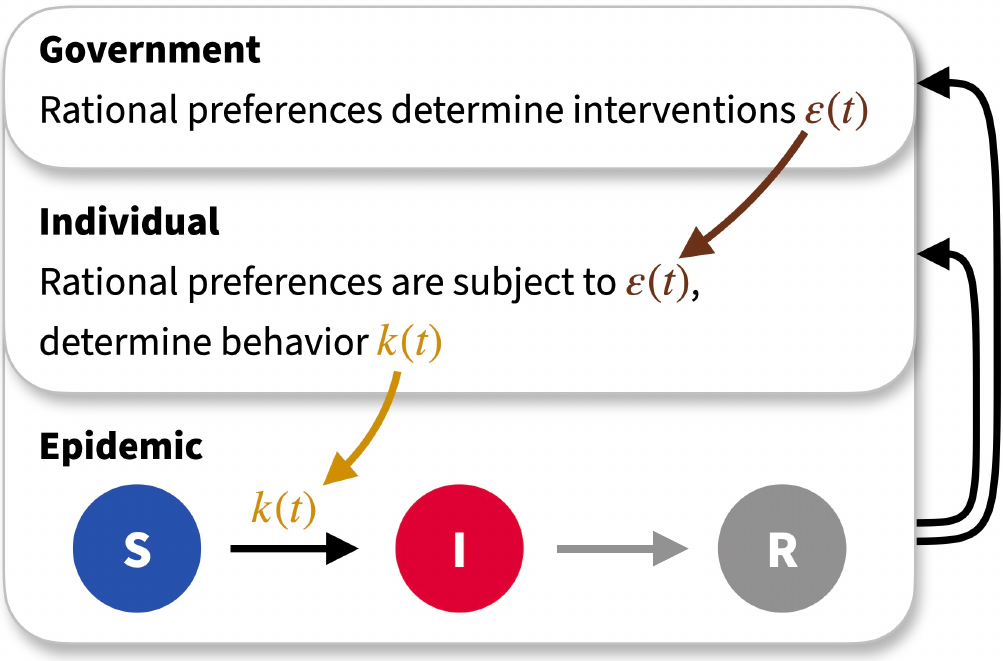}
	\caption{\textbf{Causal hierarchy of the model.} Epidemic dynamics are modelled using a simple {\underline S}usceptible-{\underline I}nfected-{\underline R}ecovered compartmental model. This informs all decision making (black arrows). The progress of the disease depends only on the behaviour of individuals, who adopt a behaviour consistent with an infectiousness $k(t)$ at time $t$ (gold arrow).  Individuals may receive government incentives $\varepsilon(t)$ (brown arrow) to modify their behaviour. They then adopt a rational strategy $k(t)$, corresponding to a Nash equilibrium, based on some utility functional. The government maximises its own value functional and intervenes with incentives for individuals to realise this. This intervention process will, in general, itself carry costs.}
	\label{fig:model_schematic}
\end{figure}

\subsection*{Epidemic dynamics}

The epidemic dynamics represent the lowest hierarchy in our problem, see Fig \ref{fig:model_schematic}, and inform all rational decisions made by the population and policy makers. We assume that the epidemic follows a standard SIR compartmentalised model \cite{Kermack1927} in which the fraction of the population in the {\underline s}usceptible, {\underline i}nfected and {\underline r}ecovered categories, the latter including any fatalities, obey the rescaled equations
\begin{align}
	\dot s & =-\kp\>s\>i\cr
	\dot i & =\kp\>s\>i-i \label{dimlessgeneral} \\
	\dot r & =i
	\nonumber
\end{align}
with initial values $s(0) = 1 - i_0$ and $i(0) = i_0$ at a time $t = 0$.
{We usually drop most functional dependencies, such as time here, for brevity.}
Here a dot denotes a time derivative and we have assumed a single timescale for recovery and the duration of infectiousness, for simplicity, measuring time $t$ in these units. {The course of the epidemic depends on the population averaged infectiousness $k(t)$, which arises from the behaviour of the whole population; as a shorthand, we directly denote $k(t)$ as behaviour. Social distancing performed by the population results in a reduction of $k$. At this level of the hierarchy, we take $k(t)$ as given, but we will calculate it self-consistently from individual behaviour in the next section.}

Since the following results do not depend on the recovered fraction of the population, we omit it in what follows. The solution of these equations is shown for constant $k = \kappa^* = 4$ in Fig \ref{fig:Nash_utilitarian_government} as a baseline for comparison to various scenarios with behavioural modification of $k$.
For this, we calculated the numerical solution of Eqs \ref{dimlessgeneral} with a standard ordinary differential equation solver implemented in the integrate.odeint function in the scipy Python package \cite{Virtanen2020}.

\subsection*{Nash equilibrium behaviour}
\label{sec:Hamiltonian_Nash}

{In the following we calculate the expected population behaviour if the population seeks out a Nash equilibrium. Conceptually, we are formulating a mean-field game \cite{Bensoussan,Carmona}, which can be solved with a standard optimal control theory approach \cite{OptimalControlBook}. Here, we are building on the work of Reluga TC and Galvani AP~\cite{Reluga2011}.}

{Any representative individual of the population is assumed to observe the course of the epidemic in the population, and select their behaviour $\kappa$ in response to it. The mean-field nature of the approach implies that the individual does not observe the behaviour of other individuals but only the averaged dynamics as described by Eqs \ref{dimlessgeneral}.
	The individual at any given time is either susceptible, infectious, or recovered, and their fate can be modelled as a series of discrete transition events between these states. In order to make the situation tractable, we calculate the expected probability $\psi_j(t)$ that the individual is in compartment $j$ at time $t$ as a continuous time Markov process. In direct analogy to the compartmental model for the epidemic in the population, we can write}
\begin{align}
	\dot \psi_s & =- \kd\psi_s i\cr
	\dot \psi_i & = \kd\psi_s i-\psi_i
	\label{dimlesspsigeneral}
\end{align}
with initial values $\psi_s(0) = s(0)$ and $\psi_i(0) = i(0)$.
These equations are similar to eqs.~\ref{dimlessgeneral} but involve the infected fraction of the population reservoir $i$, itself a solution to those equations. {The equations describe how susceptible individuals become infected by coming in contact with members of the infectious compartment of the wider population. If the individual becomes infected their behaviour is assumed not to affect the course of the epidemic itself.}
Reducing $\kd(t)$ has the effect of directly reducing the rate of change of $\psi_s$, i.e. increases the probability of remaining susceptible and lowers the probability of becoming infectious.

{Alternatively, one can interpret these equations as a compartmental model for course of the epidemic in a small group of individuals, small enough compared to the whole population so as not to affect the course of the epidemic itself, being able to employ a different strategy $\kd(t)$ as compared to the population-averaged strategy $\kp(t)$. One also has to assume that the individuals are dispersed in the population and cannot infect each other, only becoming infected by coming into contact with the rest of the population.}

{The individual knows exactly how many susceptibles, infected and recovered there are in the population, but the individual does not have any information about which group any given person belongs to. As a result, the individual cannot selectively socially distance, i.e. only distance from infected. We require everybody to socially distance.”
}

According to expected utility theory the individual will seek to maximise a utility functional which depends on both their own and the population behaviour, $U(\kappa(t), k(t))$.
{Any given individual cannot influence the behaviour of the whole population, so from the viewpoint of the individuals $k(t)$, and as a result $s(t)$ and $i(t)$ represent external or exogenous quantities, to which the individual can merely react with their own behaviour $\kappa(t)$. For the individual, the situation can be represented as a standard optimal control problem.
}

A Nash equilibrium for a population of identical individuals is found when one identifies a strategy $\kappa(t)$ for which, when adopted by the general population, individuals cannot find an alternative strategy $\tilde\kappa(t)$ that improves their utility
\begin{align}
	U(\tilde \kappa(t), \kappa(t)) \leq U(\kappa(t), \kappa(t))
	\text{, for any } \tilde\kappa(t).
\end{align}
{In such a situation, any given individual would be expected to react to the population strategy $\kappa$ by selecting behaviour $\kappa$ themselves, thus upholding the population strategy self-consistently.}

The strategy to obtain explicit solutions, is to maximise $U(\kappa, k)$ over $\kd$, treating $\kp$ as exogenous. Having identified this extremum, one sets $\kp=\kd$ to obtain the Nash equilibrium strategy adopted by the entire population. In more detail:

We analyse a simple stylised form for the individual utility with discounted utility per time $u$
\begin{align}
	U & = \int_{0}^{\infty} u(t) dt \label{U}                                                                          \\
	u & = f^{-t}\left[-\alpha(i)\> \psi_i -\beta\>(\kd -\kd^\star)^2+ (\kd -\kd^\star)\varepsilon(t)\right]\label{Uda}
\end{align}
Here $f\ge 1$ is the individual's discount rate
(equivalent to a discount time $1/\log f$). The cost associated with infection, including the risk of death, is written $\alpha(i)$. This can reflect escalating costs when a healthcare threshold is exceeded, e.g. as hospitals become full and
{as a result average treatment quality deteriorates and fatality rates increase. For simplicity we neglect the queueing process which determines whether an individual still receives state of the art healthcare such as admission to an intensive care unit with access to ventilators, etc. Instead, we assume simply that the more infectious there are on the population level, the worse on average the treatment of an individual becomes and the higher the probability of dying. Therefore the cost per single infection $\alpha(i)$ in general depends on the number of infectious $i$. 
	We study two situations. One situation is characterised by the cost of an infection being always the same, i.e. $\alpha(i) = const$. The other situation represents the fact that healthcare systems have limited capacity by having the infection cost rise near a healthcare threshold $i_{hc}$
	\begin{align}
		\alpha(i) = \alpha_0 + \frac{\alpha_1 - \alpha_0}{2}\left(\tanh[(i - i_{hc})\sigma] + 1\right)
		\label{eq:healthcare_threshold}
	\end{align}
	with minimum cost $\alpha_0$, maximum cost $\alpha_1$ and a steepness $\sigma$, see Fig \ref{fig:hct}A. If, during the course of the epidemic, the fraction of infectious $i$ approaches the threshold, the cost to being infectious increases. This reflects the greater damage from becoming infected when healthcare resources are saturated, as well as an increased likelihood of death.
}

The constant $\beta$ parameterises the financial and social costs associated with an individual modifying their behaviour from the baseline infectivity $\kappa^*$. Our choice of a quadratic form here ensures a natural equilibrium at $\kappa = \kappa^*$ in the absence of disease and/or intervention. In what follows we restrict ourselves to the case $\beta > 0$, i.e. where social distancing incurs a cost. {The edge case $\beta = 0$ changes the control problem fundamentally and leads to so-called bang-bang style behaviour $\kappa = 0$.} We can therefore
choose units for all utilities {and costs} in which $\beta = 1$ without loss of generality.

Government incentives (if any), are written $\eps(t)$.  These represent state level incentives (or penalties) designed to modify behaviour. For example, if $\eps < 0$, the government is incentivising cautious behaviour $\kappa < \kappa^*$ and taxing risky behaviour $\kappa > \kappa^*$.
The interpretation of $\eps$ as a tax/incentive would imply that whatever balance the government earns or spends by enacting $\eps$ is ultimately equally redistributed among the population.

{We have chosen a strongly idealised model and utility function, in the hope of capturing relevant behaviour without adding unnecessary model complexity. It is common to model the social activity, $\kappa$ or $k$ here, as entering linearly in the epidemic model, e.g \cite{Reluga2010,Makris2020}, and equally common to assume a convex, and in particular quadratic control cost, e.g. \cite{Makris2020,Eichenbaum2021}. Others have used different, but also convex, functional forms for the social distancing cost, e.g. \cite{Reluga2010}, and while there are some quantitative differences in the results, the Nash equilibrium behaviour is qualitatively quite similar. As we will see, the results greatly depend on whether there is a step in the infection cost or not, but we believe the particular functional shape of the step to not be relevant, qualitatively. One, however, finds very different outcomes when one assumes that the infection cost per infection \textit{decreases} with the number of infections \cite{Bhattacharyya2019}, which can result in infection waves.} We have assumed that all individuals have to pay the cost of social distancing equally
in contrast to other work, e.g. \cite{Reluga2010, Toxvaerd2020} where the cost of social distancing is paid mostly by the $s$-compartment.
Their choice is motivated by the fact that only susceptibles can influence their fate with their own behaviour.
Our choice is motivated by the observation that no individual, regardless of compartment, can socially distance without incurring a cost and
aligns more closely with the situation in which an individual doesn't necessarily know in which compartment they are, e.g. the limit where many infections occur asymptomatically.
	This would result in individuals that are in the infectious or recovered compartments acting as if they are susceptible.
	Treating this precisely would require a model with significant additional complexity.
	Future versions of our model may include explicitly for instance asymptomatic and exposed compartments, with separate controllable behaviours for each compartment.
	In addition, there can be peer-pressure effects for conformity across all compartments. 
{As for the functional choice of the government intervention: we strongly idealised the situation and assumed that the government intervention acts as a bias on the behaviour $\kappa$ linearly, with the aim of allowing the government to both incentivise more or less activity. Alternative approaches would have been for the government to use an incentive $\eps$ to influence the cost of social distancing $\beta$ by replacing the term $-\beta (\kappa-\kappa^*)^2$ with $-(\beta - \eps)(\kappa-\kappa^*)^2$ (positive $\eps<\beta$ allow $\kappa$ to deviate more easily from $\kappa^*$) or to use $\eps$ as a tax to affect the cost of an infection $\alpha(i)$ by replacing the term  $-\alpha(i)\psi_i$ by $-(\alpha(i)+\eps)\psi_i$ (positive $\eps$ encourages more social distancing to avoid the increased infection cost). These ideas would be quite similar to what is explored in Ref. \cite{Rowthorn2020}. We believe that these choices would still allow the government to target the global optimum of its objective function by appropriately incentivising/taxing the population. Since we see our work as a proof of concept, we have only focused on one type of government intervention.}

It is numerically convenient to truncate the utility integral at a final time $t_f$. Indeed this can be realistic if associated, e.g. with the rollout of mass vaccination. The contribution to the utility from the course of the epidemic after $t_f$ is written $U_f$.
Assuming the arrival of a perfect vaccine at $t_f$, which reduces the fraction of susceptibles immediately to $0$ {and thus immediately reduces the incentive to social distance, $\kappa = \kappa^*$}, the utility then reads
	\begin{align}
		U =   & \ \int_{0}^{t_f} u(t) dt + U_f  \label{eq:U_with_salvage} \\
		U_f = & \ \int_{t_f}^\infty u(t) dt
		= \int_{t_f}^\infty f^{-t}[-\alpha(i(t)) \psi_i(t)] dt
	\end{align}
	which can be numerically integrated. For convenience, we approximated the salvage term
	\begin{align}
		U_f	\approx -f^{-t_f}\alpha(0)\ \frac{\psi_{i,f}}{1+\log f}
	\end{align}
see section D in S1 Text for a short derivation.
We always choose $t_f$ large enough so that $i_f$ is extremely small (typically $\lesssim 10^{-8}$). As a result the approximation above is satisfied well and in addition $U_f$ is negligible. However, the small contribution of $U_f$ is always included in the figures and solutions we show here, for completeness. We note {again} that the arrival of a vaccine or treatment
earlier during the course of the epidemic {tends to enhance social distancing efforts \cite{Reluga2010,Makris2020,Schnyder2023}. If $\alpha(i)$ is not constant in that situation, the above approximation will not be accurate}. However, in this work, $t_f$ is assumed to always be sufficiently late for vaccination to have no behavioural or policy consequences.

{The individual behaviour $\kappa$ is assumed to satisfy the constrained optimisation problem
	\begin{align}
		\kappa      & = \argmax_\kappa U \cr
		\text{subject to }
		\dot \psi_s & =- \kd\psi_s i \text{, with }\ \psi_s(0) = 1-i_0, \text{and}\cr
		\dot \psi_i & =  \kd\psi_s i-\psi_i \text{, with }\ \psi_i(0) = i_0.
	\end{align}
	The population behaviour $k(t)$ and therefore $s(t)$ and $i(t)$, as well as the government intervention $\eps(t)$ are treated as external or exogenous quantities, outside of the individual's control. They merely represent an explicit time-dependence of the utility function and the individual's dynamics, to which the individual reacts by adjusting their behaviour without being able to affect them.}

{The solution to this optimisation problem can be calculated within a standard Hamiltonian/Lagrangian approach.
	Lev Pontryagin discovered that instead of solving the constrained optimisation problem directly, one can derive a simpler to solve set of differential equations that comprise a boundary value problem (BVP)~\cite{Pontryagin}. As an intermediate step of deriving the BVP one defines a Hamiltonian. What is now known as Pontryagin's Principle loosely states that an optimal control to solve the constrained optimisation problem must also solve this BVP, which in turn means that it also extremises the Hamiltonian.
	The BVP is equivalent to the Hamiltonian equations or Euler-Lagrange equations known in physics which can be derived when extremising an action integral.
	See sections A and B in S1 Text, or references \cite{OptimalControlBook,Reluga2011}, for a derivation and more details.
	Here we use this approach and, instead of solving the constrained optimisation problem directly, solve the BVP involving the Hamiltonian.
	The system's Hamiltonian for the individual behaviour is given by, see section B in S1 Text,
	}
\begin{align}
	H
	= & \ u + v_s (-\kd\psi_s i) + v_i (\kd\psi_s i-\psi_i) \cr
	= & -f^{-t}\left[\alpha(i) \psi_i +\betaGhost\,(\kd -\kd^\star)^2 - \varepsilon\, (\kd -\kd^\star)\right] \cr
	  & - (v_s - v_i) \kappa \psi_s i - v_i \psi_i
\end{align}
{Using this Hamiltonian, we can obtain additional differential equations and a condition on the control, which when solved together yield the optimal control.}
The Lagrange multipliers $v_s(t)$ and $v_i(t)$ constrain the dynamics to obey eqs.~(\ref{dimlesspsigeneral}). Furthermore, they can be seen as expressing the expected (economic) value of being in state $s$ and $i$, respectively, at any given time.  
The Hamiltonian equations for the values (also called costate equations in the control theory literature) are
\begin{align}
	\dot v_s & = -\frac{\partial H}{\partial \psi_s} =  (v_s - v_i) \kappa i
	\cr
	\dot v_i & = -\frac{\partial H}{\partial \psi_i} = f^{-t} \alpha(i) + v_i
	\label{eq:individual_optimal_values}
\end{align}
with boundary conditions
\begin{align}
	v_s(t_f)	= \frac{\partial U_f}{\partial \psi_{s,f}} = 0
	,\
	v_i(t_f) = \frac{\partial U_f}{\partial \psi_{i,f}} = \frac{-f^{-t_f}\alpha(0)}{1 + \log f}.
	\label{eq:individual_values_bc}
\end{align}
The Nash equilibrium strategy for an individual follows from $0 = \partial H/\partial \kappa$
and reads
\begin{align}
	\kappa = \kappa^* - \frac{f^t}{2\betaGhost} (v_s - v_i) \psi_s i + \frac{1}{2\betaGhost}\eps.
\end{align}
{as long as this expression yields a plausible, non-negative value for $\kappa$, and $\kappa = 0$, otherwise. There are some subtleties with how this bound has to be enforced during numerical solution of the equations, which we describe in section C in S1 Text.}

{Having obtained the optimal individual behaviour $\kappa$ for any given population behaviour $k$ which gives rise to the course of the epidemic $i$, we can now select the special case that constitutes a Nash equilibrium.}
Assuming that all individuals in the {population are identical and would all independently choose the same strategy in response to a given population behaviour, we can then conclude that the average behaviour of the whole population has to be identical to each individual's behaviour, thus becoming the equilibrium behaviour}, $k(t) = \kappa(t)$. Then, $s = \psi_s$ and $i = \psi_i$, as well as
{
	\begin{align}
		k = \kappa = \max\left(0, \kappa^* - \frac{f^t}{2\betaGhost} (v_s - v_i) s i + \frac{1}{2\betaGhost}\eps\right).
		\label{eq:individual_optimal_kappa}
	\end{align}
}
Therefore, we expect social distancing to increase with how strongly the state of being susceptible is valued w.r.t. the state of being infectious, and to increase with the number of susceptibles as well as the infectious.

{The equilibrium outcome of the epidemic can now easily be calculated for an exogenous government intervention field $\eps$. This is achieved by numerically solving the boundary value problem of Eqs \ref{dimlessgeneral} with boundary conditions $s(0) = 1 - i_0$ and $i(0) = i_0$, Eqs \ref{eq:individual_optimal_values} with boundary conditions Eqs \ref{eq:individual_values_bc}, in conjunction with Eq \ref{eq:individual_optimal_kappa}. We choose $i_0 = 3\cdot 10^{-8}$ and $\kappa^* = 4$ and disregard discounting, $f=1$.
	We use a typical numerical approach for such optimal control problems, a forward-backward sweep, see section C in S1 Text, or ref. \cite{OptimalControlBook} for more details and examples.
	Other methods for solving boundary value problems, such as a shooting method, would be applicable as well.}
Even though the cost of infection is a function of $i$, the objective function is convex in $\psi_i$, so we expect this optimisation problem to have a unique solution.
{As an example, this Nash solution is shown in Fig \ref{fig:Nash_utilitarian_government} for a constant infection cost $\alpha = 400$.
	The Nash behaviour leads to social distancing and therefore, compared to the non-behavioural case of $k=\kappa^*$, a longer duration for the epidemic with correspondingly lower infection rates and a smaller number of cases overall.}

\subsection*{Utilitarian maximum}

For comparison with the Nash equilibrium, we calculate the best possible population behaviour, corresponding to the limit of full cooperation on the level of individuals. {This corresponds to directly optimising the corresponding population level utility of the same form}
\begin{align}
	U_{p}   & = \int_0^{t_f} u_{p}(t)dt + U_{p,f} \cr
	u_{p}   & = f^{-t}\left[-\alpha(i)\> i -\betaGhost(\kp -\kd^\star)^2 + \varepsilon (\kp -\kd^\star)\right]\cr
	U_{p,f} & =\ -   \frac{f^{-t_f} \alpha(0) i_f}{1+\log f}
	\label{eq:Upop}
\end{align}
to find the optimal $k$, subject to Eqs \ref{dimlessgeneral} being satisfied. If adopted by the entire society this would yield the best possible outcome for all.
For convenience, we use the same variable names for the Lagrange multipliers.
{Following the formalism described in section B in S1 Text again, the corresponding Hamiltonian is}
\begin{align}
	H_{p}
	= & \ u_{p} + v_s (-k s i) + v_i (ksi - i) \cr
	= & -f^{-t}\left[\alpha(i) i +\betaGhost(\kp -\kd^\star)^2 - \varepsilon (\kp -\kd^\star)\right] - (v_s-v_i) k s i - v_i i
\end{align}
and {the Lagrange multipliers or expected values follow}
\begin{align}
	\dot v_s & = -\frac{\partial H_{p}}{\partial s} = (v_s - v_i) k i                                                                             \\
	\dot v_i & = -\frac{\partial H_{p}}{\partial i} = f^{-t}[\alpha(i) + \frac{\partial\alpha(i)}{\partial i} i] + (v_s - v_i) ks + v_i \nonumber
\end{align}
with boundary conditions
\begin{align}
	v_s(t_f) = 0,\
	v_i(t_f) = -  \frac{f^{-t_f} \alpha(0)}{1+\log f}
\end{align}
The optimal strategy follows from $0 = \partial H_p/\partial k$
and yields the same decision rule as given by Eq \ref{eq:individual_optimal_kappa} for the Nash equilibrium. {The utilitarian behaviour ends up differing from the equilibrium behaviour because the equation for the Lagrange multiplier $v_i$ gains a term $(v_s - v_i) ks$ that expresses the cost incurred from any infection causing further infections, which a self-interested individual does not consider.}
In general, the utilitarian optimum yields a higher utility than the Nash equilibrium, but is susceptible to defection by individuals who can gain at a personal level {at the expense of the rest of the population} by adopting different strategies, up to the Nash equilibrium, see Fig \ref{fig:Nash_utilitarian_government}.
{The Utilitarian behaviour can also be calculated with the forward-backward sweep method, see section C in S1 Text.}

\subsection*{Government intervention strategy}

The government's objectives are encoded in an objective function which has the same structure as the individual's but can have different parameter values
\begin{align}
	\Ug & = \int_{0}^{t_f} \ug(t) dt + \Ug_f \label{Ugova}                                                                      \\
	\ug & = f_g^{-t}\left[-\alpha_g(i) i-\beta_g\>(\kp-\kd^\star)^2-\gamma_g\>\varepsilon\>(\kp-\kd^\star)\right] \nonumber \cr
	V_f & = - \frac{f_g^{-t_f} \alpha_g(0) i_f}{1+\log f_g}.
\end{align}
with $f_g$ a governmental discount rate and where $\alpha_g$, $\beta_g$ and $\gamma_g$ account for the different costs assigned to outcomes, and interventions, at the government level.
The sign change in the intervention term means that incentivising the population can be costly to the government. The pre-factor $\gamma_g$ can account for how the cost of interventions can influence the government objective function. This is a way to model the shadow cost of public funds, i.e. the loss of utility due to the distortion of markets, etc., as caused by government intervention. The case of perfectly efficient intervention is given by $\gamma_g=0$, while $\gamma_g > 0$ implies a loss of utility due to the process of intervention itself.
We denote the lower and upper limits of $\alpha_g(i)$ as $\alpha_{g0}$ and $\alpha_{g1}$, using the same sharpness $\sigma$ as for individuals, see Eq \ref{eq:healthcare_threshold}. The small term $\Ug_f$ again models vaccination at $t_f$.

An important aspect of our work is that we investigate the situation in which the cost of an infection relative to the cost of social distancing can be quite different for the government than for an individual, $\alpha/\beta < \alpha_g/\beta_g$.
For instance, it is likely to be more difficult for an individual to negotiate the right to work remotely than if the government imposes these arrangements.

The equilibrium behaviour expressed by Eq \ref{eq:individual_optimal_kappa} uniquely determines the outcome of the epidemic in the presence of an imposed government policy $\eps(t)$. We can therefore rewrite the SIR model as a function not of $k$, but of $\eps$
\begin{align}
	\dot s & =-\kp(\eps)\>s\>i\cr
	\dot i & =\kp(\eps)\>s\>i-i
	\label{eq:sir_gov}
\end{align}
In this spirit, it is the government determining the outcome of the epidemic with its choice of $\eps$.
In analogy to individual decision making, we now have an objective function and equations for the course of the epidemic that depend on a single control variable, but instead of optimising for $\kappa$, we optimise for $\eps$.
{The complete government optimisation problem can therefore be framed as a constrained optimisation in $\eps$, $s$, and $i$, such that
	\begin{align}
		\eps   & = \argmax_{\eps} \left[\int_{0}^{t_f} \ug(t) dt + \Ug_f\right] 	\cr
		\ug(t) & = f_g^{-t}[-\alpha_g(i) i-\betagGhost (\kp(\eps)-\kd^\star)^2 -\gamma_g\>\varepsilon\>(\kp(\eps)-\kd^\star)] \cr
		V_f    & = - \frac{f_g^{-t_f} \alpha_g(0) i_f}{1+\log f_g}\cr
		\text{subject to }
		\dot s & =-\kp(\eps)\>s\>i   \text{, with }\ s(0) = 1-i_0, \text{and}  \cr
		\dot i & =\kp(\eps)\>s\>i-i  \text{, with }\ i(0) = i_0
	\end{align}
	where $k(\eps)$ is obtained from solving its own constrained optimisation problem, Eqs \ref{dimlessgeneral}, \ref{eq:individual_optimal_values}, \ref{eq:individual_values_bc}, \ref{eq:individual_optimal_kappa}, as already discussed above.
}
{We can follow the formalism described in section B in S1 Text again, noting that in the government optimisation $\eps$ now represents the control.}
The Hamiltonian for the government policy requires the introduction of two new Lagrange multipliers, $\lambda_s$ and $\lambda_i$, (dropping most functional dependencies for brevity)
\begin{align}
	H_g
	= & \ -f_g^{-t}[\alpha_g i+\betagGhost\>(\kp(\eps) -\kd^\star)^2  +\gamma_g\>\varepsilon\>(\kp(\eps) -\kd^\star)]  \cr
	  & - (\lambda_s-\lambda_i) \kp(\eps) \>s\>i  -\lambda_i i
\end{align}
Then the differential equations for the values are{, using Eq \ref{eq:individual_optimal_kappa}}
\begin{align}
	\dot \lambda_s = & \ -\frac{\partial H_g}{\partial s}
	= i \Lambda
	\cr
	\dot \lambda_i = & \ -\frac{\partial H_g}{\partial i}
	= s \Lambda + f_g^{-t}\left[\alpha_g(i) + \alpha_g'(i)i\right] + \lambda_i
	\label{eq:government_optimal_values}
	\\
	\Lambda =        & \  - (\lambda_i-\lambda_s) \left(\kappa^\star + \frac{\epsilon}{2 \betaGhost} \right) + \frac{f_g^{-t}f^t}{2} (v_i-v_s) \times \cr
	                 & \ \left(
	i s\left[\betagGhost f^{ t} (v_i-v_s) -2 f_g^t (\lambda_i-\lambda_s) \right]
	+  \epsilon  (\betaGhost  \gamma_g + \betagGhost)
	\right)
	\nonumber
\end{align}
with boundary conditions
\begin{align}
	\lambda_s(t_f)
	= 0,\
	\lambda_i(t_f)
	= -   \frac{f_g^{-t_f} \alpha_g(0)}{1+\log f_g}
	\label{eq:government_optimal_values_bc}
\end{align}

The optimal government strategy obeys $0 =\partial H_g/\partial \eps$ which yields{, using Eq \ref{eq:individual_optimal_kappa}}
\begin{align}
	\eps & = is \frac{f^t(\betagGhost + \gamma_g)(v_s - v_i)-f_g^t\betaGhost(\lambda_s - \lambda_i)}{\betagGhost + 2 \gamma_g}
	\label{eq:government_optimal_intervention}
\end{align}

{We can obtain the government strategy with a nested application of a forward-backward sweep of Eqs \ref{eq:sir_gov}, \ref{eq:government_optimal_values}, \ref{eq:government_optimal_values_bc}, \ref{eq:government_optimal_intervention}, see section C in S1 Text. At each iteration, we use the current estimate of the optimal government strategy $\eps$ to calculate the Nash equilibrium behaviour $k(\eps)$, also with a forward-backward sweep and as described aboved, as part of the forward integration of the dynamics. This secondary forward-backward sweep treats government intervention as exogenous. 
}

{In the case of a constant infection cost, the government's objective function is convex and we expect numerically obtained solutions to be unique. In the case of a healthcare threshold, the government's objective function is not convex, in contrast to the individuals' objective function. For that situation} it is therefore not straight-forward to establish uniqueness of our numerical solutions to the optimisation problem. In fact, by varying the initial guesses for the controls in the nested optimisation, we always found  exactly two local optima for each set of parameters -- never more or less -- and selected the one with higher utility. We take this as indication that we successfully identify the global maximum in each case.

\begin{figure}[tbph!]
	\centering \includegraphics[width=.7\columnwidth]{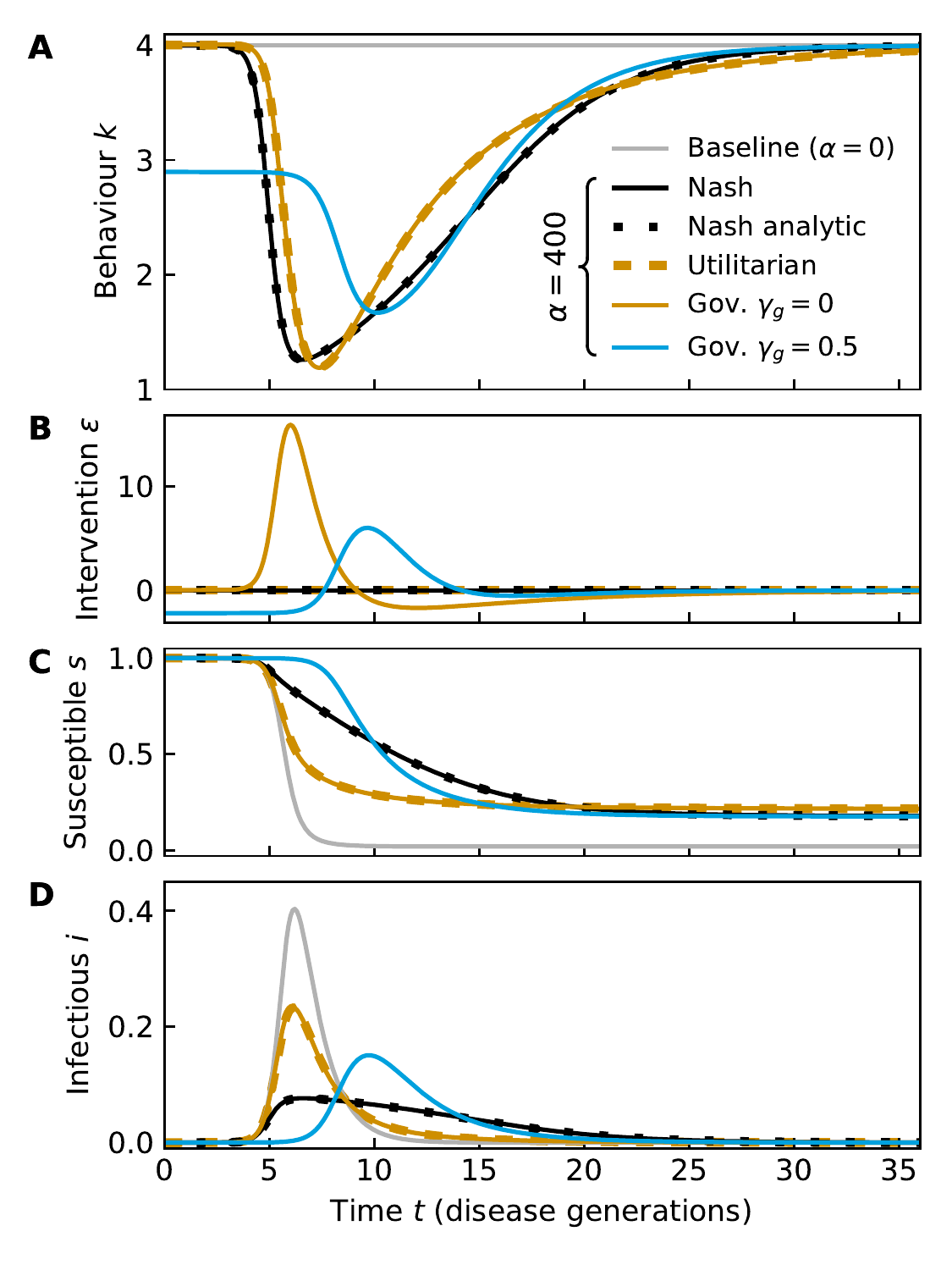}
	\caption{
		{\bf Comparison of social distancing behaviour.}
		{(A) Population behaviour $k(t)$, (B) government intervention $\eps(t)$ and (C, D) dynamics of the disease $s$, $i$ for a range of scenarios with $i_0=3\cdot 10^{-8}$, $f=1$, and $\kappa^* = 4$ throughout:
			a baseline where there is no behavioural modification (corresponding to equilibrium behaviour at an infection cost $\alpha=0$, grey lines);
			the Nash equilibrium for $\alpha=400$ (black lines), calculated numerically via forward-backward sweep, see section C in S1 Text (In order to demonstrate that the numerical solution is accurate, we also show the analytical solution of the same equations \cite{SchnyderTurnerAnalytic} as black dots); the utilitarian maximum for $\alpha=400$ (gold dashes); and finally the population behaviour for two optimal government policies, one being without cost to the government, $\gamma_g = 0$ (gold lines), and one being costly, $\gamma_g = 0.5$, with $\alpha=400$ (cyan lines). When government interventions are cost-free, they enable the population to reach the utilitarian maximum.}
	}
	\label{fig:Nash_utilitarian_government}
\end{figure}

{As an example, we calculate the government intervention strategy and the resulting incentivised equilibrium strategy for the situation where the government and individuals share the same preferences, $\alpha_g = \alpha = 400$, $\betagGhost = \beta = 1$. If government intervention is free of cost for the government, $\gamma_g = 0$, the optimal government strategy $\eps(t)$ targets the utilitarian maximum for the population, {see the gold lines in} Fig \ref{fig:Nash_utilitarian_government}. To achieve the utilitarian maximum, the $\eps$ field is used to bias the individuals' equilibrium strategy away from the unperturbed Nash equilibrium to coincide with the utilitarian maximum.
}

\section*{Results}

{Even though we strove for simplicity in our modeling choices, the model has a great number of parameters, $\kappa^*, f, \alpha_0, \alpha_1, \alpha_{g0}, \alpha_{g1}, i_{hc}, \sigma, \beta, \beta_g, \gamma_g$. We are therefore working in a moderately high-dimensional parameter space which would be challenging to fully explore. For simplicity, we adopted values representative of a disease like Covid-19. We selected single values for $\kappa^*, f, \alpha_0, \sigma, \beta, \beta_g$, while focusing on the effects of varying $\alpha_1,\alpha_{g1}, i_{hc}, \gamma_g$ to study the full range of behaviours and incentive strategies that might be expected to occur.}

\subsection*{Results without government intervention}

At first, we concentrate on the case where the cost of infection is constant, {$\alpha(i) = const$} and where the government takes no role in the response to the epidemic, $\eps = 0$. This situation has been already discussed for slightly different utilities, e.g. \cite{Reluga2010}.
To appreciate the impact of optimal decision making, it is helpful to first establish a baseline: the course of an epidemic without any behavioural modification, $k = \kappa^*$, see Fig \ref{fig:Nash_utilitarian_government} (grey curves). This corresponds to a situation where there is no perceived risk associated with an infection, $\alpha = 0$.
{Since there is no behavioural modification, Fig \ref{fig:Nash_utilitarian_government}A, and no government intervention, Fig \ref{fig:Nash_utilitarian_government}B, the number of susceptibles $s$ quickly drops, Fig \ref{fig:Nash_utilitarian_government}C, as they become infected, Fig \ref{fig:Nash_utilitarian_government}D. The peak of $i$ is extremely large, with a fraction of roughly $0.4$ of the whole population being infected at once. Consequently, the fraction of people that remains uninfected at the end of the epidemic, $s_\infty$, which is shorthand for $s(t\to\infty)$, reaches close to 0, Fig \ref{fig:Nash_utilitarian_government}C.
	In contrast, for a moderate risk, $\alpha = 400$ (black lines), the population chooses a Nash equilibrium with considerable reduction in their activity $k$, Fig \ref{fig:Nash_utilitarian_government}A, which reduces peak infection levels, Fig \ref{fig:Nash_utilitarian_government}D, and the total number of cases $1-s_\infty$, Fig \ref{fig:Nash_utilitarian_government}C, at the expense of prolonging the epidemic. }
The utilitarian optimum {(gold lines)} can target a scenario where the duration of the epidemic is almost the same as for the baseline scenario, with a smaller total of cases than for the Nash equilibrium.

\begin{figure}[tbph!]
	\begin{adjustwidth}{-2.25in}{0in} 
		\centering \includegraphics[width=7.3in]{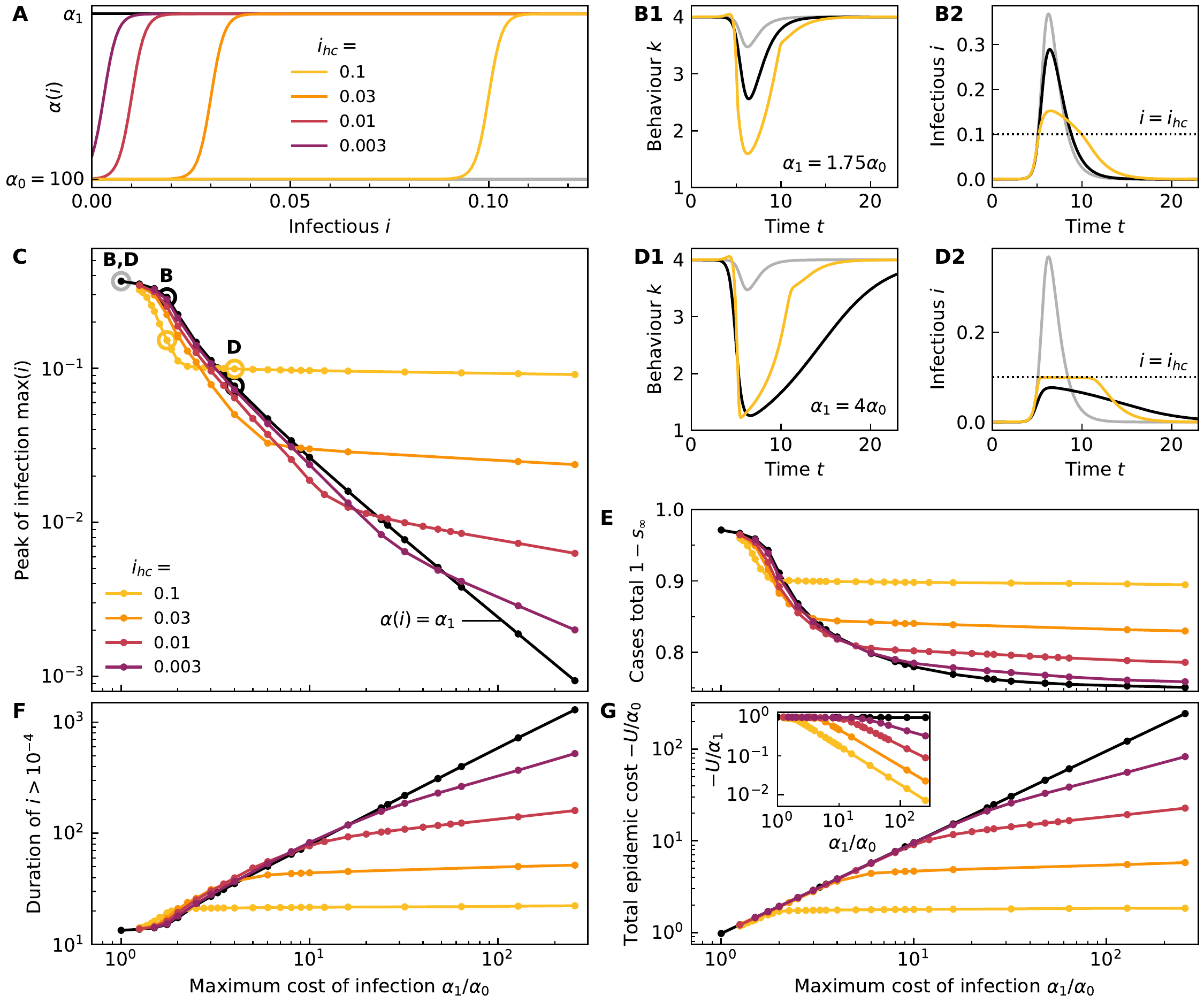}
		\caption{\textbf{The rational behaviour in the presence of a healthcare threshold depends on the maximum cost of infection.}
			{(A) The infection cost $\alpha(i)$ for a range of healthcare thresholds $i_{hc}$, see Eq \ref{eq:healthcare_threshold} with steepness $\sigma=300$. The colours encode the position of the healthcare threshold $i_{hc}$ for the whole figure.
				For comparison, two scenarios where $\alpha(i) = \alpha_0$ (grey line) and $\alpha(i) = \alpha_1$ (black) are considered as well. The base infection cost is kept constant throughout, $\alpha_0 = 100$, whereas $\alpha_1$ is varied in the following panels.
				(B) Typical example for the equilibrium behaviour of population $k(t)$ (B1) and the corresponding infectious cases $i$ (B2) over time for low maximum infection cost $\alpha_1$. We compare the behaviour for a  healthcare threshold, $\alpha_0 = 100$, $\alpha_1= 1.75\alpha_2$ with $i_{hc} = 0.1$ (yellow) to the behaviour for constant infection costs $\alpha(i) = 100$ (grey) and $\alpha(i) = 1.75\alpha_0$ (black) These cases are also marked in panel (C) by correspondingly coloured circles beneath the letters B.
				(C) The peak of the epidemic $\max(i)$ as a function of the maximum cost of being infectious $\alpha_1$ corresponding to the infection cost scenarios shown in (A). We also mark the data points corresponding to the examples shown in panels (B1-2) and (D1-2) with circles and corresponding labels above.
				(D) Typical example for the equilibrium behaviour of population $k(t)$ (D1) and the corresponding infectious cases $i$ (D2) over time for high maximum infection cost $\alpha_1$. We compare the behaviour for a healthcare threshold, $\alpha_0 = 100$, $\alpha_1= 4\alpha_2$ with $i_{hc} = 0.1$ (yellow) to the behaviour for constant infection costs $\alpha(i) = 100$ (grey) and $\alpha(i) = 4\alpha_0$ (black) These cases are also marked in panel (C) by correspondingly coloured circles beneath the letters D.
				For the same data as in panel (C):
				(E) Total number of cases after the epidemic has run its course.
				(F) Duration of the epidemic as defined by the time interval for which $i > 10^{-4}$.
				(G) Total cost of the epidemic $-U$ for equilibrium behaviour in units of the minimal infection cost $\alpha_0$. In the inset, the epidemic cost is shown in units of the maximum infection cost.
				Lines in (D-G) serve as guides to the eye.}
		}
		\label{fig:hct}
	\end{adjustwidth}
\end{figure}

The higher the cost of infection $\alpha$, the stronger is the behavioural modification, see black lines in Fig \ref{fig:hct}; see also examples for the equilibrium behaviour for $\alpha=100$ (grey lines) and $\alpha=175$ (black lines) in Fig \ref{fig:hct}B1 and \ref{fig:hct}B2, as well as $\alpha=400$ (black lines) in Fig \ref{fig:hct}D1 and \ref{fig:hct}D2.
As a consequence, the peak height of infections, Fig \ref{fig:hct}C, and the total number of cases, Fig \ref{fig:hct}E, are reduced with increasing $\alpha$, whereas the epidemic duration, Fig \ref{fig:hct}F, increases. The duration on which the behaviour deviates from the pre-epidemic value $\kappa^*$ is comparable to the duration of the epidemic.
Even though the total number of cases decreases with rising $\alpha$, the total epidemic cost $-U$ grows approximately proportionally to $\alpha$, Fig \ref{fig:hct}G, see in particular the inset where the black line is almost exactly a constant. This implies that the gains in utility by avoiding cases in excess of herd immunity are almost completely offset by the cost of social distancing.

Next, we express the fact that healthcare systems have limited capacity by having the infection cost rise near a healthcare threshold $i_{hc}$, see Eq \ref{eq:healthcare_threshold}.
{
We investigate the outcomes for a number of thresholds, see Fig \ref{fig:hct}A. We vary the value of $i_{hc}$ while keeping the absolute steepness of the transition $\sigma=300$ constant. This has the effect that the relative steepness $\sigma i_{hc}$ varies with $i_{hc}$, with the transitions being the steeper, the larger the threshold. This enables us to investigate the effects of threshold location and transition steepness at the same time.
We set $\alpha_0 = 100$ and vary $\alpha_1$ in relation to that.
In passing, we note that $\alpha(i)$ is a monotonically increasing function and that the cost per infection at $i=0$ is not necessarily exactly $\alpha_0$, but
$\alpha(0)= \frac{1}{2}(\alpha_0 + \alpha_1 - (\alpha_1-\alpha_0)\tanh(i_{hc} \sigma)) > \alpha_0$. For the healthcare thresholds that we studied, the difference can be completely neglected for $i_{hc} = 0.1$ and $0.03$, whereas for $i_{hc}=0.01$ one obtains a correction of $\alpha(0) \approx  \alpha_0  + 2.5\times 10^{-3}(\alpha_1 - \alpha_0)$ and for $i_{hc}=0.003$, $\alpha(0) \approx  \alpha_0  + 0.14(\alpha_1 - \alpha_0)$.
	}

Varying the maximum infection cost $\alpha_1$ at a given threshold $i_{hc}$, we find in general two qualitatively different Nash equilibrium strategies, see Fig \ref{fig:hct}B-\ref{fig:hct}G.
For instance, let us focus on $i_{hc}=0.1$ for now, see the yellow lines.

	{
		(1)~Low infection cost strategy: For low $\alpha_1$, it is rational to enact stronger social distancing than for the case of a constant high cost of infection, $\alpha = \alpha_1$.
		As an illustrative example, we show the equilibrium behaviour in the situation where the infection cost rises from $\alpha_0=100$ to $\alpha_1 = 1.75\alpha_1$ at $i_{hc}=0.1$ in Fig \ref{fig:hct}B1 and \ref{fig:hct}B2 and compare that with the limiting cases of having constant infection cost $\alpha=\alpha_0$ (grey lines) and $\alpha=\alpha_1$ (black lines).
		We find that social distancing in the presence of the threshold is stronger than for both constant cost cases. It is obvious that social distancing would be more extreme when there is a healthcare threshold at which the cost increases from $\alpha = \alpha_0$ to $\alpha_1$ than if $\alpha=\alpha_0$ always. But it is perhaps surprising that the situation with a healthcare threshold would call for stronger social distancing as compared to the case where $\alpha=\alpha_1$ always, given that the time averaged infection cost in the presence of the threshold is lower without any additional social distancing. However, the additional investment in social distancing is more than offset by the reduction in infection cost.
		Still, the peak of infection generally exceeds the health care threshold, see Fig \ref{fig:hct}D, if only slightly in the example.
		Strategy (1) is found to the left of the constant infection cost line for $\alpha=\alpha_1$ (black) in Fig \ref{fig:hct}C.
		Strategy (1) is also characterised by lowered case numbers as compared to the constant infection cost case, Fig \ref{fig:hct}E and slightly longer epidemic durations, Fig \ref{fig:hct}F.
		The total epidemic cost $-U$ is only slightly lower than for constant infection cost, in fact it is almost imperceivable on the scale of Fig \ref{fig:hct}G.
		Focusing again on Fig \ref{fig:hct}C, we see that the higher the infection cost, the lower the infection peak becomes, until it approximately meets the health care threshold at $\alpha_1 \approx 2\alpha_0-3\alpha_0$ for $i_{hc}=0.1$, where the situation crosses over into:}

(2)~High infection cost strategy: If $\alpha_1$ exceeds a critical value which depends on $i_{hc}$ (and to a lesser extent on $\alpha_0$ and $\sigma$), the
rational strategy is not to exceed the health care threshold but to remain close to it. An illustrative example for the equilibrium behaviour in the situation where the infection cost rises from $\alpha_0=100$ to $\alpha_1 = 4\alpha_1$ at $i_{hc}=0.1$ is shown as yellow line in Fig \ref{fig:hct}D1 and \ref{fig:hct}D2, comparing to the limiting cases of having constant infection cost $\alpha=\alpha_0$ (grey lines) and $\alpha=\alpha_1=4\alpha_1$ (black lines). This strategy yield less severe social distancing than the constant $\alpha = \alpha_1$ case. As a result we observe higher peaks of infection, Fig \ref{fig:hct}C, with a higher total of cases, Fig \ref{fig:hct}E, and shorter duration of the epidemic, Fig \ref{fig:hct}F, when compared to the constant infection cost case. However, since the healthcare threshold is generally not exceeded, the total epidemic cost is much lower than in the constant infection cost case, Fig \ref{fig:hct}G.

For the lower healthcare thresholds (darker colours), we find qualitatively similar behaviour. However, the more slowly $\alpha(i)$ varies at the threshold,  Fig \ref{fig:hct}A, the more gradual is the transition between strategies (1) and (2). For the lower values of $i_{hc}$, e.g. $i_{hc}=0.003$ (purple lines), the peak of infection keeps decreasing with increasing $\alpha_1$ across the whole studied range, Fig \ref{fig:hct}C. This is due to the fact that the infection cost $\alpha(i)$ does not reach the constant value $\alpha_0$ for finite $i$ and thus any reduction in $\max(i)$ can yield a lower infection cost.
For larger $i_{hc}$, $\max(i)$ becomes practically independent of $\alpha_1$ at large $\alpha_1$ because the infection cost at the peak has already reached $\alpha(\max(i)) = \alpha_0$. Corresponding tendencies are found for the total cases, epidemic duration, as well as total epidemic cost.

Fig \ref{fig:hct}G shows, that if $\alpha_1$ is held constant, the total epidemic cost strongly decreases with increasing $i_{hc}$.
This underlines the potentially significant benefit of investing in healthcare infrastructure in order to raise $i_{hc}$.

	\subsection*{Results with government intervention}

	If government and individuals share the same preferences, $\alpha_g(i) = \alpha(i)$, $\betagGhost = \beta = 1$,
	and if government intervention is free of cost for the government, $\gamma_g = 0$, the optimal government strategy $\eps(t)$ gives rise to the utilitarian maximum for the population, {see the gold lines in} Fig \ref{fig:Nash_utilitarian_government} for an example where the infection cost is constant. To achieve the utilitarian maximum, the $\eps$ field is used to bias the individuals' equilibrium strategy in the presence of government intervention away from the unperturbed Nash equilibrium to coincide with the utilitarian maximum.
	{If the government wishes to encourage more cautious behaviour, it selects $\eps < 0$, which rewards behaviour $\kappa < \kappa^*$ and taxes $\kappa > \kappa^*$.}
	Owing to the level of control the government has over the population with its intervention strategy, the government is able to achieve a course of the epidemic that is shorter while resulting in fewer infections in total. It achieves this by initially incentivising social activity and later on incentivising social distancing in a precisely controlled manner. It is very encouraging that this closely resembles the strategy of the Japanese government, with its ``Go To campaign'' from July 2020 onwards. This was designed to increase demand for domestic tourism. This was eventually phased out and replaced with policies to promote social distancing.

	We note that when $\gamma_g = 0$, individual preferences are irrelevant for the course of the epidemic: The government will always be able to find an intervention strategy $\eps$ that makes the population's equilibrium behaviour align with the government's preferences. However, the greater the difference in preferences, the greater the amplitude of $\eps$ necessary to achieve this.

	Next, we consider the case when intervention is costly for the government, see the cyan lines in Fig \ref{fig:Nash_utilitarian_government} for an example where the infection cost is constant and $\gamma_g = 0.5$. then we find that the government selects an intervention strategy , Fig \ref{fig:Nash_utilitarian_government}B, which incentivises population behaviour that is less strongly varying over time, Fig \ref{fig:Nash_utilitarian_government}A. The government does not necessarily intervene less, but it chooses to incentivise social distancing earlier in time so that the peak of social distancing can be less extreme, Fig \ref{fig:Nash_utilitarian_government}A. Note that this policy yields fewer uninfected at the end of the epidemic, $s(t\to\infty)$, Fig \ref{fig:Nash_utilitarian_government}C, as the socially optimal strategy. This occurs even though the peak of the epidemic is lower Fig \ref{fig:Nash_utilitarian_government}D.

\begin{figure}[tbp]
		\begin{adjustwidth}{-2.25in}{0in} 
			\centering \includegraphics[width=7.4in]{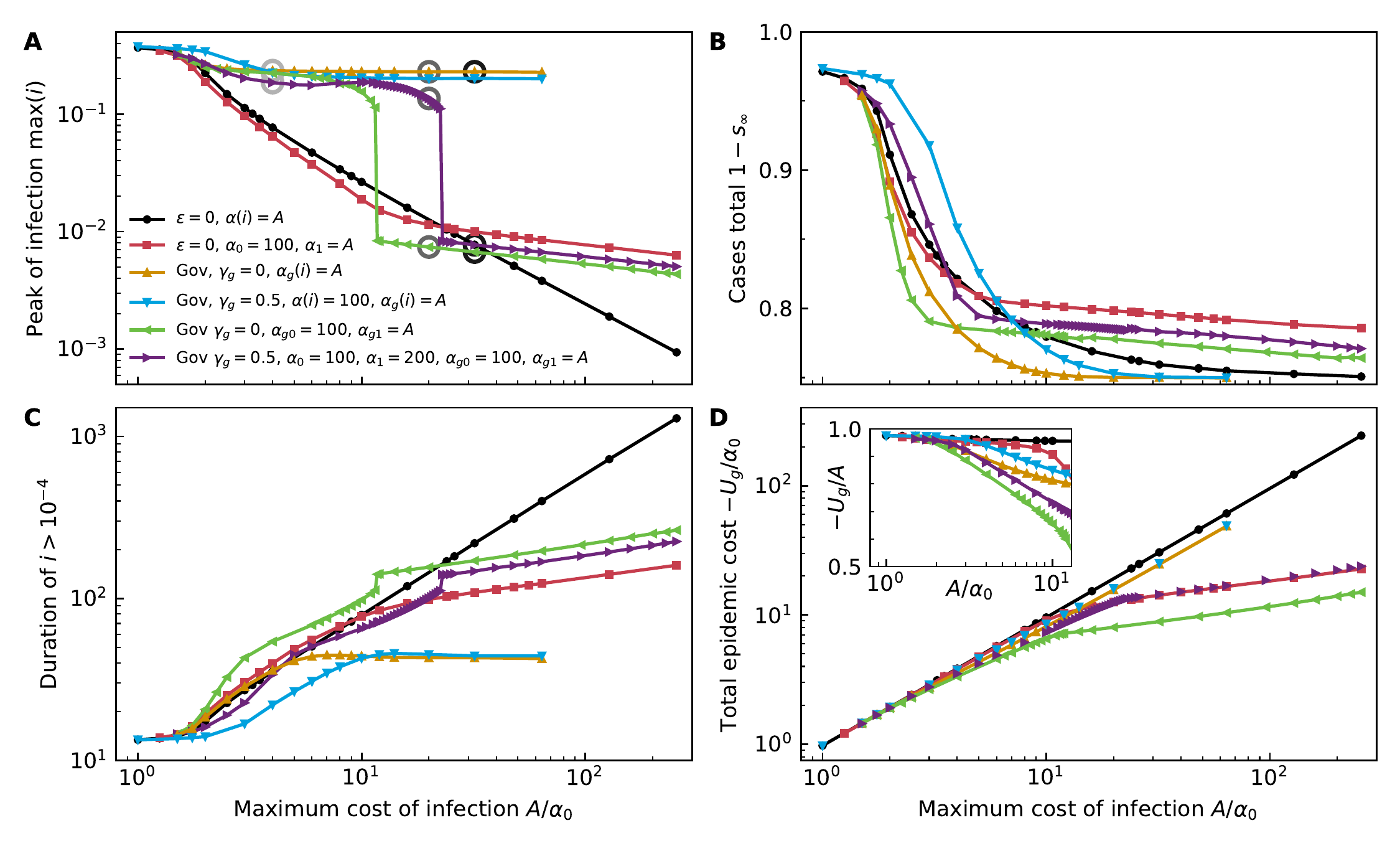}
			\caption{\textbf{Optimal government policy.}
				{(A) Peak of the epidemic as a function of the maximum cost of infection for a range of scenarios where the (maximum) infection cost for either population or government is varied:
					Nash equilibrium behaviour of the population, without government intervention for a constant infection cost (black line, replotted from Fig \ref{fig:hct}) and with a healthcare threshold at $i_{hc}=0.01$ (red, replotted from Fig \ref{fig:hct});
					with government intervention for a constant infection cost $\alpha_g$ (cost-free $\gamma_g=0$: gold, costly $\gamma_g = 0.5$: cyan)
					and with a healthcare threshold at $i_{hc}=0.01$ (cost-free $\gamma_g=0$: green,  costly $\gamma_g = 0.5$: purple). The circles mark the scenarios shown in Fig \ref{fig:hct_gov_2}. For these scenarios, we also show
					(B) the total number of cases,
					(C) the duration of the epidemic, as well as
					(D) the total cost of the epidemic in units of the minimal infection cost $\alpha_0$. In the cases without government intervention, the total cost is calculated as $-U$, whereas in the cases with government intervention, we report $-U_g$. In the inset, the epidemic cost is shown in units of the maximum infection cost. Lines serve as guides to the eye.}
			}
			\label{fig:hct_gov}
		\end{adjustwidth}
	\end{figure}

\begin{figure}[tbp]
		\begin{adjustwidth}{-2.25in}{0in} 
			\centering \includegraphics[width=7.5in]{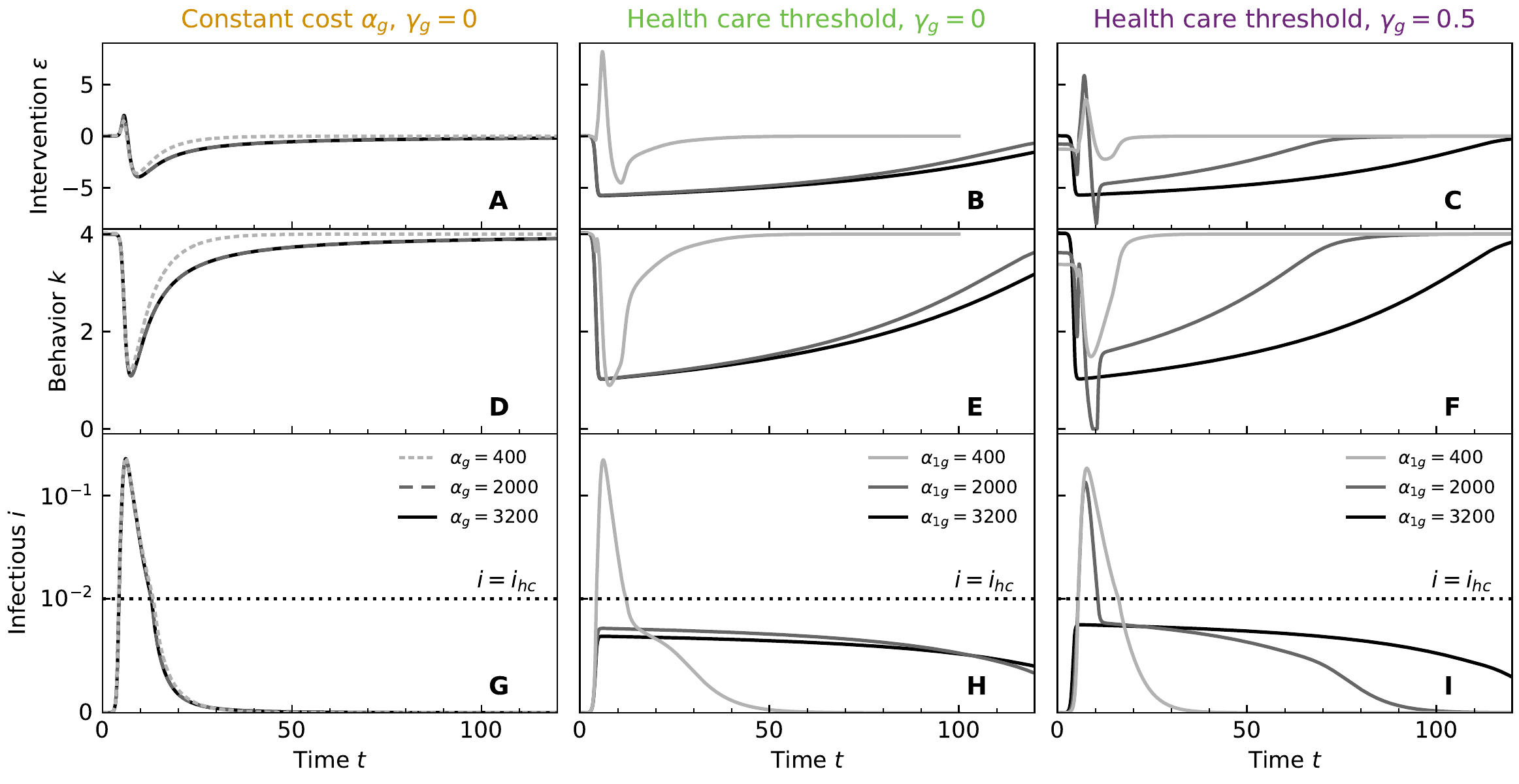}
			\caption{\textbf{Course of the epidemic with government intervention.}
				{Government intervention $\eps$ assuming (A) constant infection cost $\alpha_g$ and cost-free intervention $\gamma_g=0$, (B) a healthcare threshold (HT) and cost-free intervention $\gamma_g=0$, and (C) a healthcare threshold and costly intervention $\gamma_g = 0.5$; all for a range of $a_{g}$ or $a_{1g}$, respectively, as marked by circles in Fig \ref{fig:hct_gov}A and listed in the legends of (G-I), with individuals assuming that $\alpha_0 = 100$.
					(D-F) Equilibrium population behaviour $k$ in response to $\eps$ of (A-C), respectively.
					(G-I) Infectious $i$ over time, corresponding to the behaviour shown in (D-F), respectively. Here, the $y$-axis has linear scale between $0$ and $10^{-2}$ and logarithmic scale above that.}
			}
			\label{fig:hct_gov_2}
		\end{adjustwidth}
	\end{figure}

	For constant infection cost, the government strategy only weakly depends on the infection cost, regardless of whether the intervention is costly or not: The peak of infection is relatively insensitive to $\alpha$ for both cost-free and costly intervention, see gold and cyan lines in Fig \ref{fig:hct_gov}A respectively. However, the total number of cases approaches the herd immunity threshold more rapidly with intervention than without (black line), Fig \ref{fig:hct_gov}B. Cost-free intervention enables this at lower infection cost than costly intervention. Government intervention also manages to keep the duration of the epidemic much shorter than without incentives, Fig \ref{fig:hct_gov}C, at a lower total epidemic cost, Fig \ref{fig:hct_gov}D. The inset, in which the total epidemic cost is normalised by the maximum infection cost, shows this more clearly.
		The intervention policy, Fig \ref{fig:hct_gov_2}A, and its effect on population behaviour, Fig \ref{fig:hct_gov_2}D, and the course of the epidemic, Fig \ref{fig:hct_gov_2}G, vary only subtly with rising infection cost. The larger the infection cost, the longer social distancing is practised and the more gradually it is relaxed over extended periods of time.

	If the capacity of the healthcare system is limited according to
	Eq \ref{eq:healthcare_threshold}, see Fig \ref{fig:hct}A, government intervention leads to a markedly different course of the epidemic as compared to {the no-intervention equilibrium, see green and purple lines in Fig \ref{fig:hct_gov} and compare with the red lines. (We show data with $i_{hc}=0.01$ but expect the scenario to be qualitatively the same for other thresholds.)
		Instead of a continuous reduction of the infection peak without government intervention, Fig \ref{fig:hct_gov}A, incentives lead to a} sharp switch between policies that favour high peak infection and those that track the health care threshold as the maximum infection cost $\alpha_{1g}$ increases.
	At low $\alpha_{1g}$ the government targets a solution with a higher peak of infections, Fig \ref{fig:hct_gov}A, without necessarily increasing the total number of cases, Fig \ref{fig:hct_gov}B, at the expense of a longer duration of the epidemic, Fig \ref{fig:hct_gov}C. This strategy resembles the government strategy for constant infection cost.
	When the maximum infection cost is high, the government targets the healthcare threshold, albeit at a lower peak of infections than the population would be able to reach on its own. The crossover between the regimes depends on the direct cost for the interventions, controlled by the parameter $\gamma_g$.  When the intervention is cost-free $\gamma_g = 0$, the crossover occurs at a markedly smaller maximum infection cost than for the case without government intervention.

	{Cost-free intervention enables a significantly lower total epidemic cost than no intervention,Fig \ref{fig:hct_gov}D, as it targets the utilitarian optimum. As compared to no intervention, the costly intervention scenario also results in lowered total epidemic cost at low maximum infection cost but roughly the same total cost at high maximum infection costs. However, it achieves this by lowering the total case numbers which is offset by the cost of intervening.}

	For reference, we show the government intervention, the behaviour of the population in response to it, and the course of the epidemic for a range of maximum infection costs $\alpha_{1g}$ in Fig \ref{fig:hct_gov_2}.

	Regarding the switch in the government strategy {, which leads to the sharp jump in the infection peak observed in Fig \ref{fig:hct_gov}A, and examples of which we show in Fig \ref{fig:hct_gov_2}B and \ref{fig:hct_gov_2}C}: As stated earlier, we always find two locally optimal solutions. These form branches, with one being globally optimal for low maximum infection cost and one being optimal for large maximum infection cost.
	{We only show the solutions that are globally optimal, but the two branches both appear as linear on the log-log plot of Fig \ref{fig:hct_gov}D, one at low infection cost and one at large infection cost.}
	The switch in the government strategy {occurs at the $\alpha_{g1}$ at which} these branches yield the same value of the objective function.
	The policy that is optimal at low maximum infection cost $\alpha_{g1}$ is characterised by a high infection peak{, Fig \ref{fig:hct_gov}A,} and shorter epidemic duration{, Fig \ref{fig:hct_gov}C}; it therefore tolerates higher infection numbers in order to reduce costs incurred from social distancing. The policy that is optimal at large maximum infection cost $\alpha_{g1}$ favours an investment in stronger social distancing to avoid infections.
	The policy under high/low infection costs results in a greater/lower $s_\infty${, Fig \ref{fig:hct_gov}B}.
	While the crossover between these policies is continuous in the maximum values of the objective function{, Fig \ref{fig:hct_gov}D}, it results in very different disease trajectories, in particular in a discontinuous change in the peak infections{, Fig \ref{fig:hct_gov}A}.
In contrast, the Nash solution in the absence of government control is smooth because the ability of individuals to defect from an optimal consensus strategy leads to a smoothing out of the switch{, Fig \ref{fig:hct_gov}A}.

	\section*{Conclusion and discussion}

	{Here, we have shown how costly interventions, such as taxes or subsidies on behaviour, can be used to exactly align individuals' decision making with government preferences even when these are not aligned. In order to achieve this, we developed a nested optimisation algorithm of both the government intervention strategy and the resulting equilibrium behaviour of individuals.}
	Healthcare systems in general, and intensive care facilities in particular, have limited capacity. For instance, intensive care units in Japan, the UK, and Germany had approximately 5, 7, and 34 beds per 100,000 people, respectively, in April 2020~\cite{Nikkei,Guardian}, with most of them regularly occupied. Assuming a {healthcare} threshold above which costs rise as a result of the rationing of scarce (intensive) care resources among patients, we find that it can be rational to adjust behaviour so that infections remain close to this threshold. This is a generic response when {\it either} the above-threshold costs $\alpha_1$ or $\alpha_{1g}$, for the individuals or government respectively, are high enough.
	However, the disease dynamics can be very different under government intervention than without it, see e.g. Fig \ref{fig:hct_gov}A. We find that optimal government intervention strategies undergo a sharp ``switch'' from high peak infection numbers to a lower level, around the healthcare threshold. Furthermore, we find that both the maximum infection cost at which this switch occurs and the form of the intervention adopted are sensitive to how costly the intervention is to the government. For diseases that have infection costs around the value at which this policy switch occurs we anticipate that it would be very difficult for policymakers to know whether to adopt a high- or low-peak infection approach, particularly in the face of uncertainties. In the context of the COVID-19 epidemic it may be that the costs {were} such that the system {was} located close to this switch. This might help to explain why government policies to tackle COVID-19 differed so markedly between countries. Crude back-of-the-envelope analysis indicates that this may indeed be the case, although we are reluctant to assign values, this being fundamentally a political decision. In particular, if the infection cost were an order of magnitude higher/lower, policy determination would be straightforward.

Our results also show that a dramatic reduction in total epidemic cost can be achieved by increasing the healthcare threshold, implying the policy recommendation to do so.

Future work could include expanding our formalism to noisy dynamics, noisy control~\cite{yong1999stochastic,Lorch2018,Tottori2022,Tottori2023b,Tottori2023}, imperfect information or to study the robustness of the control, similar to \cite{Kantner2020,Kohler2021,Morris2021}. There is also the intriguing possibility of allowing individuals to directly influence government \cite{Mellacher2023}
	in the same way that $\eps$ allows the government to influence individuals. One approach might be to model political contentment, controlled by individuals, that would appear in the government objective function. This could give rise to a formalism with significant game theoretic complexity.

\section*{Acknowledgments}
	
	We would like to dedicate this work to the memory of Prof. George Rowlands, who passed away in early 2021 and who was involved in many of the early discussions leading to this work.
	We would like to thank Andrew J. Oswald, Steven Satchell, Michael Tildesley, Giorgos Galanis, Patrick Mellacher, and Holly C. Turner for careful reading of the manuscript and helpful comments. We also thank Tetsuya J. Kobayashi and Shuhei A. Horiguchi for helpful discussions.
	Matthew S. Turner acknowledges the kind hospitality of the Yamamoto group.

	The numerical code was written in Python, in particular using the numpy \cite{Harris2020} and scipy \cite{Virtanen2020} packages. Figures were generated using the Matplotlib Python library \cite{Hunter2007}.


\begin{thebibliography}{}

\bibitem{Reluga2010}
Reluga TC.
\newblock {Game Theory of Social Distancing in Response to an Epidemic}.
\newblock PLoS Comput Biol. 2010;6(5):e1000793.
\newblock doi:{10.1371/journal.pcbi.1000793}.

\bibitem{Fenichel2011}
Fenichel EP, Castillo-Chavez C, Ceddia MG, Chowell G, Parra PAG, Hickling GJ,
  et~al.
\newblock {Adaptive human behavior in epidemiological models}.
\newblock \new{Proceedings of the National Academy of Sciences of the United States of America.} 2011;108(15):6306--6311.
\newblock doi:{10.1073/pnas.1011250108}.

\bibitem{Wang2016}
Wang Z, Bauch CT, Bhattacharyya S, D'Onofrio A, Manfredi P, Perc M, et~al.
\newblock {Statistical physics of vaccination}.
\newblock Physics Reports. 2016;664:1--113.
\newblock doi:{10.1016/j.physrep.2016.10.006}.

\bibitem{Chang2020}
Chang SL, Piraveenan M, Pattison P, Prokopenko M.
\newblock {Game theoretic modelling of infectious disease dynamics and
  intervention methods: a review}.
\newblock Journal of Biological Dynamics. 2020;14(1):57--89.
\newblock doi:{10.1080/17513758.2020.1720322}.

\bibitem{Verelst2016}
Verelst F, Willem L, Beutels P.
\newblock {Behavioural change models for infectious disease transmission: A
  systematic review (2010-2015)}.
\newblock Journal of the Royal Society Interface. 2016;13(125).
\newblock doi:{10.1098/rsif.2016.0820}.

\bibitem{Bhattacharyya2019}
Bhattacharyya S, Reluga T.
\newblock {Game dynamic model of social distancing while cost of infection
  varies with epidemic burden}.
\newblock IMA J Appl Math (Institute Math Its Appl. 2019;84(1):23--43.
\newblock doi:{10.1093/imamat/hxy047}.

\bibitem{Makris2020}
Makris M, Toxvaerd F.
\newblock {Great Expectations : Social Distancing in Anticipation of
  Pharmaceutical Innovations}; 2020. 2097.

\bibitem{Yan2021}
Yan Y, Malik AA, Bayham J, Fenichel EP, Couzens C, Omer SB.
\newblock {Measuring voluntary and policy-induced social distancing behavior
  during the COVID-19 pandemic}.
\newblock Proceedings of the National Academy of Sciences of the United States
  of America. 2021;118(16):1--9.
\newblock doi:{10.1073/pnas.2008814118}.

\bibitem{Reluga2011}
Reluga TC, Galvani AP.
\newblock {A general approach for population games with application to
  vaccination}.
\newblock Math Biosci. 2011;230(2):67--78.
\newblock doi:{10.1016/j.mbs.2011.01.003}.

\bibitem{Acemoglu2020}
Acemoglu D, Chernozhukov V, Werning I, Whinston MD.
\newblock {Optimal targeted lockdowns in a multi-group SIR model}.
\newblock National Bureau of Economic Research; 2020. 27102.

\bibitem{Prem2017}
Prem K, Cook AR, Jit M.
\newblock {Projecting social contact matrices in 152 countries using contact
  surveys and demographic data}.
\newblock PLOS Computational Biology. 2017;13(9):e1005697.
\newblock doi:{10.1371/journal.pcbi.1005697}.

\bibitem{Huang2022}
Huang CI, Crump RE, Brown PE, Spencer SEF, Miaka EM, Shampa C, et~al.
\newblock {Identifying regions for enhanced control of gambiense sleeping
  sickness in the Democratic Republic of Congo}.
\newblock Nature Communications. 2022;13(1):1--11.
\newblock doi:{10.1038/s41467-022-29192-w}.

\bibitem{Tildesley2022}
Tildesley MJ, Vassall A, Riley S, Jit M, Sandmann F, Hill EM, et~al.
\newblock {Optimal health and economic impact of non-pharmaceutical
  intervention measures prior and post vaccination in England: a mathematical
  modelling study}.
\newblock Royal Society Open Science. 2022;9(8).
\newblock doi:{10.1098/rsos.211746}.

\bibitem{Chandrasekhar2021}
Chandrasekhar AG, Goldsmith-Pinkham P, Jackson MO, Thau S.
\newblock {Interacting regional policies in containing a disease}.
\newblock Proceedings of the National Academy of Sciences of the United States
  of America. 2021;118(19):1--7.
\newblock doi:{10.1073/pnas.2021520118}.

\bibitem{Holme2012}
Holme P, Saram{\"{a}}ki J.
\newblock {Temporal networks}.
\newblock Physics Reports. 2012;519(3):97--125.
\newblock doi:{10.1016/j.physrep.2012.03.001}.

\bibitem{Holme2015}
Holme P, Masuda N.
\newblock {The basic reproduction number as a predictor for epidemic outbreaks
  in temporal networks}.
\newblock PLoS ONE. 2015;10(3):1--15.
\newblock doi:{10.1371/journal.pone.0120567}.

\bibitem{He2013}
He D, Dushoff J, Day T, Ma J, Earn DJD.
\newblock {Inferring the causes of the three waves of the 1918 influenza
  pandemic in England and Wales}.
\newblock Proceedings of the Royal Society B: Biological Sciences.
  2013;280(1766).
\newblock doi:{10.1098/rspb.2013.1345}.

\bibitem{Mossong2008}
Mossong J, Hens N, Jit M, Beutels P, Auranen K, Mikolajczyk R, et~al.
\newblock {Social contacts and mixing patterns relevant to the spread of
  infectious diseases}.
\newblock PLoS Medicine. 2008;5(3):0381--0391.
\newblock doi:{10.1371/journal.pmed.0050074}.

\bibitem{Tildesley2010}
Tildesley MJ, House TA, Bruhn MC, Curry RJ, O'Neil M, Allpress JLE, et~al.
\newblock {Impact of spatial clustering on disease transmission and optimal
  control}.
\newblock Proceedings of the National Academy of Sciences of the United States
  of America. 2010;107(3):1041--1046.
\newblock doi:{10.1073/pnas.0909047107}.

\bibitem{Sun2021}
Sun K, Wang W, Gao L, Wang Y, Luo K, Ren L, et~al.
\newblock {Transmission heterogeneities, kinetics, and controllability of
  SARS-CoV-2}.
\newblock Science. 2021;371(6526):eabe2424.
\newblock doi:{10.1126/science.abe2424}.

\bibitem{Ferguson2006}
Ferguson NM, Cummings DAT, Fraser C, Cajka JC, Cooley PC, Burke DS.
\newblock {Strategies for mitigating an influenza pandemic}.
\newblock Nature. 2006;442(7101):448--452.
\newblock doi:{10.1038/nature04795}.

\bibitem{Tanimoto2018}
Tanimoto J.
\newblock {Social Dilemma Analysis of the Spread of Infectious Disease}.
\newblock Springer Nature Singapore; 2018.

\bibitem{Mellacher2020}
Mellacher P.
\newblock {COVID-Town: An Integrated Economic-Epidemiological Agent-Based
  Model}.
\newblock Graz Schumpeter Centre; 2020. 23.

\bibitem{Grauer2020}
Grauer J, L{\"{o}}wen H, Liebchen B.
\newblock {Strategic spatiotemporal vaccine distribution increases the survival
  rate in an infectious disease like Covid-19}.
\newblock Scientific Reports. 2020;10(1):1--10.
\newblock doi:{10.1038/s41598-020-78447-3}.

\bibitem{yong1999stochastic}
Yong J, Zhou XY.
\newblock Stochastic controls: Hamiltonian systems and HJB equations. vol.~43.
\newblock Springer Science \& Business Media; 1999.

\bibitem{Molina2022}
Molina JJ, Schnyder SK, Turner MS, Yamamoto R.
\newblock {Nash Neural Networks : Inferring Utilities from Optimal Behaviour}.
  2022;.
\newblock \new{arXiv:2203.13432.}

\bibitem{kahneman2003maps}
Kahneman D.
\newblock Maps of bounded rationality: Psychology for behavioral economics.
\newblock American economic review. 2003;93(5):1449--1475.

\bibitem{McAdams2020}
McAdams D.
\newblock {Nash SIR: An Economic-Epidemiological Model of Strategic Behavior
  During a Viral Epidemic}.
\newblock Covid Economics. 2020;doi:{10.2139/ssrn.3593272}.

\bibitem{Eichenbaum2021}
Eichenbaum MS, Rebelo S, Trabandt M.
\newblock {The Macroeconomics of Epidemics}.
\newblock Review of Financial Studies. 2021;34(11):5149--5187.
\newblock doi:{10.1093/rfs/hhab040}.

\bibitem{Rowthorn2020}
Rowthorn R, Toxvaerd F.
\newblock {The optimal control of infectious diseases via prevention and
  treatment}.
\newblock University of Cambridge; 2020. 2027.

\bibitem{Toxvaerd2020}
Toxvaerd F, Rowthorn R.
\newblock {On the management of population immunity}.
\newblock University of Cambridge; 2020. 2080.

\bibitem{Li2017}
Li J, Lindberg DV, Smith RA, Reluga TC.
\newblock {Provisioning of Public Health Can Be Designed to Anticipate Public
  Policy Responses}.
\newblock Bull Math Biol. 2017;79(1):163--190.
\newblock doi:{10.1007/s11538-016-0231-8}.

\bibitem{Bethune2020}
Bethune ZA, Korinek A.
\newblock {COVID-19 infection externalities: trading off lives vs.
  livelihoods}.
\newblock Cambridge, MA: National Bureau of Economic Research; 2020. 27009.

\bibitem{Aurell2022}
Aurell A, Carmona R, Dayanikli G, Lauri{\`{e}}re M.
\newblock {Optimal Incentives to Mitigate Epidemics: A Stackelberg Mean Field
  Game Approach}.
\newblock SIAM Journal on Control and Optimization. 2022;60(2):S294--S322.
\newblock doi:{10.1137/20M1377862}.

\bibitem{Althouse2010}
Althouse BM, Bergstrom TC, Bergstrom CT.
\newblock {A public choice framework for controlling transmissible and evolving
  diseases}.
\newblock Proceedings of the National Academy of Sciences of the United States
  of America. 2010;107(SUPPL. 1):1696--1701.
\newblock doi:{10.1073/pnas.0906078107}.

\bibitem{Stiglitz1971}
Stiglitz J, Dasgupta P.
\newblock Differential Taxation, Public Goods, and Economic Efficiency.
\newblock Review of Economic Studies. 1971;38(2):151--174.

\bibitem{Palmer2020}
Palmer AZ, Zabinsky ZB, Liu S.
\newblock {Optimal control of COVID-19 infection rate considering social
  costs}. 2020; p. 1--27.

\bibitem{Hayhoe2020}
Hayhoe M, Barreras F, Preciado VM.
\newblock {Data-Driven Control of the COVID-19 Outbreak via Non-Pharmaceutical
  Interventions: A Geometric Programming Approach}.
\newblock Annual reviews in control. 2020;52(NA):495--507.
\newblock doi:{10.1016/j.arcontrol.2021.04.014}.

\bibitem{Piguillem2020}
Piguillem F, Shi L.
\newblock {Optimal COVID-19 Quarantine and Testing Policies}; 2020. 20/04.

\bibitem{Kantner2020}
Kantner M, Koprucki T.
\newblock {Beyond just ``flattening the curve'': Optimal control of epidemics
  with purely non-pharmaceutical interventions}.
\newblock Journal of Mathematics in Industry. 2020;10(1).
\newblock doi:{10.1186/s13362-020-00091-3}.

\bibitem{Ketcheson2021}
Ketcheson DI.
\newblock {Optimal control of an SIR epidemic through finite-time
  non-pharmaceutical intervention}.
\newblock Journal of mathematical biology. 2021;83(1):7.
\newblock doi:{10.1007/s00285-021-01628-9}.

\bibitem{Schwarzendahl2022}
Schwarzendahl FJ, Grauer J, Liebchen B, L{\"{o}}wen H.
\newblock {Mutation induced infection waves in diseases like COVID-19}.
\newblock Scientific Reports. 2022;12(1):1--11.
\newblock doi:{10.1038/s41598-022-13137-w}.

\bibitem{Giannitsarou2021}
Giannitsarou C, Kissler S, Toxvaerd F.
\newblock {Waning Immunity and the Second Wave: Some Projections for
  SARS-CoV-2}.
\newblock American Economic Review: Insights. 2021;3(3):321--338.
\newblock doi:{10.1257/aeri.20200343}.

\bibitem{Lux2021}
Lux T.
\newblock {The social dynamics of COVID-19}.
\newblock Physica A: Statistical Mechanics and its Applications.
  2021;567:125710.
\newblock doi:{10.1016/j.physa.2020.125710}.

\bibitem{Bauch2003}
Bauch CT, Galvani AP, Earn DJD.
\newblock {Group interest versus self-interest in smallpox vaccination policy}.
\newblock Proceedings of the National Academy of Sciences of the United States
  of America. 2003;100(18):10564--10567.
\newblock doi:{10.1073/pnas.1731324100}.

\bibitem{Bauch2004}
Bauch CT, Earn DJD.
\newblock {Vaccination and the theory of games}.
\newblock Proceedings of the National Academy of Sciences of the United States
  of America. 2004;101(36):13391--13394.
\newblock doi:{10.1073/pnas.0403823101}.

\bibitem{Reluga2006}
Reluga TC, Bauch CT, Galvani AP.
\newblock {Evolving public perceptions and stability in vaccine uptake}.
\newblock Mathematical Biosciences. 2006;204(2):185--198.
\newblock doi:{10.1016/j.mbs.2006.08.015}.

\bibitem{Tildesley2006}
Tildesley MJ, Savill NJ, Shaw DJ, Deardon R, Brooks SP, Woolhouse MEJ, et~al.
\newblock {Optimal reactive vaccination strategies for a foot-and-mouth
  outbreak in the UK}.
\newblock Nature. 2006;440(7080):83--86.
\newblock doi:{10.1038/nature04324}.

\bibitem{Chen2014}
Chen F, Toxvaerd F.
\newblock {The economics of vaccination}.
\newblock Journal of Theoretical Biology. 2014;363:105--117.
\newblock doi:{10.1016/j.jtbi.2014.08.003}.

\bibitem{Moore2021b}
Moore S, Hill EM, Dyson L, Tildesley MJ, Keeling MJ.
\newblock {Modelling optimal vaccination strategy for SARS-CoV-2 in the UK}.
\newblock PLoS Computational Biology. 2021;17(5):1--20.
\newblock doi:{10.1371/journal.pcbi.1008849}.

\bibitem{Moore2022}
Moore S, Hill EM, Dyson L, Tildesley MJ, Keeling MJ.
\newblock {Retrospectively modeling the effects of increased global vaccine
  sharing on the COVID-19 pandemic}.
\newblock Nature Medicine. 2022;28(11):2416--2423.
\newblock doi:{10.1038/s41591-022-02064-y}.

\bibitem{Hill2022}
Hill EM, Prosser NS, Ferguson E, Kaler J, Green MJ, Keeling MJ, et~al.
\newblock {Modelling livestock infectious disease control policy under
  differing social perspectives on vaccination behaviour}.
\newblock PLOS Computational Biology. 2022;18(7):e1010235.
\newblock doi:{10.1371/journal.pcbi.1010235}.

\bibitem{Keeling2023}
Keeling MJ, Moore S, Penman BS, Hill EM.
\newblock {The impacts of SARS-CoV-2 vaccine dose separation and targeting on
  the COVID-19 epidemic in England}.
\newblock Nature Communications. 2023;14(1):1--10.
\newblock doi:{10.1038/s41467-023-35943-0}.

\bibitem{Kucharski2020}
Kucharski AJ, Klepac P, Conlan AJK, Kissler SM, Tang ML, Fry H, et~al.
\newblock {Effectiveness of isolation, testing, contact tracing, and physical
  distancing on reducing transmission of SARS-CoV-2 in different settings: a
  mathematical modelling study}.
\newblock The Lancet Infectious Diseases. 2020;20(10):1151--1160.
\newblock doi:{10.1016/S1473-3099(20)30457-6}.

\bibitem{Schnyder2023}
Schnyder SK, Molina JJ, Yamamoto R, Turner MS.
\newblock {Rational social distancing in epidemics with uncertain vaccination
  timing}.
\newblock PLOS ONE. 2023;18(7):e0288963.
\newblock doi:{10.1371/journal.pone.0288963}.

\bibitem{Kermack1927}
Kermack WO, McKendrick AG.
\newblock {A contribution to the mathematical theory of epidemics}.
\newblock Proceedings of the Royal Society of London Series A, Containing
  Papers of a Mathematical and Physical Character. 1927;115(772):700--721.
\newblock doi:{10.1098/rspa.1927.0118}.

\bibitem{Virtanen2020}
Virtanen P, Gommers R, Oliphant TE, Haberland M, Reddy T, Cournapeau D, et~al.
\newblock {SciPy 1.0: fundamental algorithms for scientific computing in
  Python}.
\newblock Nature Methods. 2020;17(3):261--272.
\newblock doi:{10.1038/s41592-019-0686-2}.

\bibitem{Bensoussan}
Bensoussan A, Frehse J, Yam P.
\newblock Mean Field Games and Mean Field Type Control Theory.
\newblock Springer; 2013.

\bibitem{Carmona}
Carmona R, Delarue F.
\newblock Probabilistic Theory of Mean Field Games with Applications I.
\newblock Springer; 2018.

\bibitem{OptimalControlBook}
Lenhart S, Workman J.
\newblock Optimal Control Applied to Biological Models.
\newblock Chapman and Hall/CRC; 2007.

\bibitem{Pontryagin}
Pontryagin LS, Boltyanskii VG, Gamkrelidze RV, Mishchenko EF.
\newblock {The Mathematical Theory of Optimal Processes}.
\newblock Gordon and Breach Science Publishers; 1986.

\bibitem{SchnyderTurnerAnalytic}
Schnyder SK, Molina JJ, Yamamoto R, Turner MS.
\newblock \new{In preparation}.

\bibitem{Nikkei}
Maemura A. Japan ranks below Italy and Spain in ICU bed capacity; 2020.
\newblock
  \url{https://asia.nikkei.com/Spotlight/Coronavirus/Japan-ranks-below-Italy-and-Spain-in-ICU-bed-capacity}.

\bibitem{Guardian}
Campbell D. NHS hospital bosses urge ministers to increase ICU beds in England;
  2021.
\newblock
  \url{https://www.theguardian.com/society/2021/feb/28/uk-government-must-increase-number-of-nhs-beds-hospital-bosses-warn}.

\bibitem{Lorch2018}
Lorch L, De A, Bhatt S, Trouleau W, Upadhyay U, Gomez-Rodriguez M.
\newblock {Stochastic Optimal Control of Epidemic Processes in Networks}.
  2018;.

\bibitem{Tottori2022}
Tottori T, Kobayashi TJ.
\newblock {Memory-Limited Partially Observable Stochastic Control and Its
  Mean-Field Control Approach}.
\newblock Entropy. 2022;24(11):1--27.
\newblock doi:{10.3390/e24111599}.

\bibitem{Tottori2023b}
Tottori T, Kobayashi TJ.
\newblock {Forward-Backward Sweep Method for the System of HJB-FP Equations in
  Memory-Limited Partially Observable Stochastic Control}.
\newblock Entropy. 2023;25(2).
\newblock doi:{10.3390/e25020208}.

\bibitem{Tottori2023}
Tottori T, Kobayashi TJ.
\newblock {Decentralized Stochastic Control with Finite-Dimensional Memories: A
  Memory Limitation Approach}.
\newblock Entropy. 2023;25(5):1--26.
\newblock doi:{10.3390/e25050791}.

\bibitem{Kohler2021}
K{\"{o}}hler J, Schwenkel L, Koch A, Berberich J, Pauli P, Allg{\"{o}}wer F.
\newblock {Robust and optimal predictive control of the COVID-19 outbreak}.
\newblock Annual Reviews in Control. 2021;51(November 2020):525--539.
\newblock doi:{10.1016/j.arcontrol.2020.11.002}.

\bibitem{Morris2021}
Morris DH, Rossine FW, Plotkin JB, Levin SA.
\newblock {Optimal, near-optimal, and robust epidemic control}.
\newblock Communications Physics. 2021;4(1):1--8.
\newblock doi:{10.1038/s42005-021-00570-y}.

\bibitem{Mellacher2023}
Mellacher P.
\newblock {The impact of corona populism: Empirical evidence from Austria and
  theory}.
\newblock Journal of Economic Behavior \& Organization. 2023;209:113--140.
\newblock doi:{10.1016/j.jebo.2023.02.021}.

\bibitem{Harris2020}
Harris CR, Millman KJ, van~der Walt SJ, Gommers R, Virtanen P, Cournapeau D,
  et~al.
\newblock {Array programming with NumPy}.
\newblock Nature. 2020;585(7825):357--362.
\newblock doi:{10.1038/s41586-020-2649-2}.

\bibitem{Hunter2007}
Hunter JD.
\newblock {Matplotlib: A 2D Graphics Environment}.
\newblock Computing in Science \& Engineering. 2007;9(3):90--95.
\newblock doi:{10.1109/MCSE.2007.55}.

\end{thebibliography}

\nolinenumbers

\clearpage

\appendix

\section*{Supporting Information}

\paragraph*{S1 Text.} This text provides a background on the calculus of variations, optimal control theory, the forward-backward sweep for solving optimal control problems, and how to calculate our utility salvage term.

\end{document}